\documentclass[prb,amsmath,amssymb,nofootinbib,10pt,notitlepage,twocolumn]{revtex4-2}

\usepackage[utf8]{inputenc}

\usepackage{graphicx}
\usepackage[dvipsnames]{xcolor}
\usepackage{hyperref}
\usepackage{braket}
\usepackage{relsize}
\usepackage{moresize}

\newcommand{\dd}{\operatorname{d\!}{}}
\newcommand{\thin}{\mkern1.5mu}

\newcommand{\V}[1]{\boldsymbol{#1}}  
\newcommand{\T}[1]{\boldsymbol{#1}}  
\newcommand{\M}[1]{\hat{#1}}  

\newcommand{\F}[1]{\mathcal{#1}} 
\newcommand{\FM}[1]{\widehat{\mathcal{#1}}}  

\newcommand{\C}{\mathcal{C}}
\newcommand{\MG}{\M G}
\newcommand{\MGs}{\M G^{\rm b}}
\newcommand{\Gs}{G^{\rm b}}

\newcommand{\tk}{\,\tilde{\!\varkappa}} 
\newcommand{\tl}{\tilde{\lambda}}
\newcommand{\tm}{\tilde{\mu}}

\DeclareMathSymbol{:}{\mathbin}{operators}{"3A} 

\newcommand{\p}{{\vphantom{()}}}
\newcommand{\Tr}{\operatorname{Tr}}
\renewcommand{\div}{\operatorname{div}}
\newcommand{\rot}{\operatorname{rot}}

\usepackage{textgreek}
\newcommand{\LC}{\text{\straightepsilon}} 

\newcommand{\Vr}{\V{r}}
\newcommand{\Vq}{\V{q}}
\newcommand{\Vp}{\V{p}}

\usepackage{mathrsfs} 
\usepackage{common-unicode} 
\newcommand{\avg}[1]{\left<#1\right>}
\newcommand{\avgs}[1]{\bigl<#1\bigr>}

\newcommand{\spar}[1]{{\text{\ssmall(}#1\text{\ssmall)}}}

\begin{document}

\title{Large-scale exponential correlations of nonaffine elastic response of strongly disordered materials}
\date{\today}
\author{D.\,A.~Conyuh}
\author{D.\,V.~Babin}
\author{I.\,O.~Raikov}
\author{Y.\,M.~Beltukov}
\date{\today}
\email{yaroslav.beltukov@mail.ioffe.ru}
\affiliation{Ioffe Institute, Politechnicheskaya st.~26, 194021 St.~Petersburg, Russia}

\begin{abstract}
    The correlation properties of the nonaffine elastic response in strongly disordered materials are investigated using the theory of correlated random matrices and supported by numerical models. While the nonaffine displacement field itself predominantly exhibits power-law decay, we demonstrate that its spatial derivatives reveal large-scale exponentially decaying correlations. Specifically, the correlation functions of the divergence and (for most deformations) the rotor of the nonaffine field are governed by a heterogeneity length scale $\xi$. This length scale is set by the disorder strength and can become indefinitely large, far exceeding the structural correlation length. A notable exception occurs under volumetric deformation, where the rotor correlations 
    lack the exponential tail with the length scale $ξ$. The theory also predicts that the rotor correlations may have small power-law tails. We directly observe the exponential decay, characterized by $\xi$, in numerical studies of a rigidity percolation model and in molecular dynamics simulations of amorphous polystyrene and the Lennard-Jones glass. The latter example also confirms the existence of the power-law tail in the rotor correlation function at large distances.
\end{abstract}

\maketitle

\section{Introduction}
\label{sec:Intro}

Interest in studying the nature of the amorphous solid state and its microscopic structure has not waned over the decades \cite{Anderson-glass-lightly-1995, Ngai-why-glass-transition-2007, McKenna-50th-anniversary-perspective-2017}. The properties of vibrational excitations and the local elastic properties of amorphous materials are the subject of active research \cite{Zhou-thermal-conductivity-amorphous-2020, Hu-origin-boson-peak-2022}. The random atomic arrangement in amorphous (glassy) materials significantly influences their behavior on both the nanoscale and macroscopic levels, with microscopic elastic properties displaying noticeable spatial inhomogeneity and diverging considerably from those found in crystalline materials \cite{Yoshimoto-mechanical-heterogeneities-model-2004, Tsamados-local-elasticity-map-2009, Wagner-local-elastic-properties-2011, Mizuno-measuring-spatial-distribution-2013}. Such properties encompass the phenomenon of \emph{nonaffine} atomic displacements.

In a perfectly ordered structure, such as Bravais lattices, a uniform macroscopic deformation leads to an affine displacement, with the displacement of each atom proportional to its position. In this case, the forces acting on a single atom from its nearest neighbors can cancel each other out due to symmetry, and there are no resulting local forces, only macroscopic ones~\cite{Zaccone-network-disorder-nonaffine-2011}. In contrast, in a disordered structure, there are nonzero local forces acting on each atom, causing additional local displacements called nonaffine~\cite{Alexander-amorphous-solids-their-1998, Lemaitre-sum-rules-quasistatic-2006, Zaccone-network-disorder-nonaffine-2011}. The presence of nonaffine displacements has been observed in a wide range of amorphous materials: metallic glasses~\cite{Jana-correlations-nonaffine-displacements-2019}, polymer hydrogels~\cite{Wen-nonaffine-deformations-polymer-2012}, supercooled liquids~\cite{DelGado-nonaffine-deformation-inherent-2008}, Lennard-Jones glasses~\cite{Goldenberg-particle-displacements-elastic-2007}, silica glass~\cite{Leonforte-inhomogeneous-elastic-response-2006}. It has been shown that nonaffine deformations are responsible for viscoelasticity~\cite{Lemaitre-sum-rules-quasistatic-2006}, internal damping~\cite{Damart-theory-harmonic-dissipation-2017}, and sound attenuation~\cite{Szamel-microscopic-analysis-sound-2022,
Szamel-sound-attenuation-glasses-2025} in amorphous solids. The suppression of nonaffine displacements leads to an enhancement of the local elastic moduli around nanoparticles of the host amorphous medium~\cite{Beltukov-local-elastic-properties-2022,Conyuh-effective-elastic-moduli-2023}.

The theoretical description of nonaffine displacements is significantly complicated by the fact that continuum elasticity theory is not applicable at scales where nonaffine deformations play a significant role~\cite{Tanguy-continuum-limit-amorphous-2002}. As a result, the heterogeneity length scale $ξ$ can be introduced, which separates large scales, where the continuum theory of elasticity applies, and microscopic scales with significant nonaffine deformations. The heterogeneity length scale depends on the strength of the disorder in the system and can be up to tens of interatomic distances~\cite{Leonforte-continuum-limit-amorphous-2005}.

At present, the theory of nonaffine displacements based on fluctuations of elastic moduli in the framework of the continuum elasticity theory is well developed. DiDonna and Lubensky~\cite{DiDonna-nonaffine-correlations-random-2005} found analytically and numerically that the nonaffine correlation function $⟨\V u^{\rm naff}(\V r)·\V u^{\rm naff}(0)⟩$ has power-law decay $r^{-1}$ for systems with dimension $d=3$, and logarithmic decay for $d=2$. Such a decay was confirmed by Maloney \emph{et al.}~\cite{Maloney-correlations-elastic-response-2006, Maloney-anisotropic-power-law-2009} while examining the elastic response of two-dimensional Lennard-Jones glasses using computer simulations. The power-law correlations were also attributed to the elastic deformation field due to plastic events, which was substantiated by numerical simulations on hard-sphere glasses~\cite{Varnik-correlations-plasticity-sheared-2014}.

On the other hand, there is evidence that nonaffine correlations decay exponentially in the elastic regime. 
Lerner and Bouchbinder~\cite{Lerner-anomalous-linear-elasticity-2023} mentioned the existence of long-range displacements correlations $\sim r^{(2-d)}\exp(-r/ξ)$ for $r<ξ$ with exponential decay at the heterogeneity length scale $ξ$, and found for $r\llξ$ an anomalous power law decay $\sim 1/r$ in response functions to a local force dipole.
Jana and Pastewka~\cite{Jana-correlations-nonaffine-displacements-2019} studied correlations of nonaffine displacements during simple shear deformation of Cu-Zr bulk metallic glasses in molecular dynamics calculations. Their calculations show an exponential correlation with a decay length, which they interpret as the size of a shear transformation zone in the elastic regime. Meenakshi and Gupta~\cite{Meenakshi-characteristics-correlations-nonaffine-2022} demonstrate that the correlation function changes from exponential to power-law decay at the yielding transition.
The reported strain magnitude in these papers can lead to plastic events, making the observed response not purely elastic, which stimulates interest in a detailed investigation of this issue in the case of a purely elastic response of the material.
Therefore, the debate about the power-law or exponential decay of the correlations of nonaffine displacement fields has been complicated by observations of both types of decay.

Thus, the development of the theory of nonaffine deformations is of great interest. The most important aspect is to take into account the heterogeneity length scale and to develop a theory that would be applicable on scales smaller than the heterogeneity length scale, which can reveal the exponential behavior of nonaffine deformations. At the heterogeneity length scale, the fluctuations of the elastic moduli are comparable to the mean elastic moduli. This not only prevents the application of continuum elasticity theory, but also raises an important issue about the mechanical stability of the disordered system. When a system is cooled from the melt to temperatures well below the glass transition temperature, it reaches a metastable equilibrium, settling into one of many local minima of potential energy, thereby becoming an amorphous solid~\cite{Stillinger-supercooled-liquids-glass-1988, Lemaitre-sum-rules-quasistatic-2006}. Therefore, the disorder in an amorphous solid is constrained by the fact that such systems are stable, implying significant correlations of the force constants~\cite{Alexander-amorphous-solids-their-1998, Lerner-frustrationinduced-internal-stresses-2018}. It has been shown that the theory of positive-definite correlated random matrices is an effective way of accounting for the strong disorder that provides mechanical stability~\cite{Beltukov-iofferegel-criterion-diffusion-2013, Conyuh-random-matrix-approach-2021}.

Random matrix theory has been applied to study the mechanical properties of amorphous glassy materials and amorphous polymers~\cite{Grigera-vibrations-glasses-euclidean-2002, Beltukov-iofferegel-criterion-diffusion-2013, Manning-random-matrix-definition-2015, Baggioli-vibrational-density-states-2019, Conyuh-random-matrix-approach-2021, Conyuh-effective-elastic-moduli-2023}. Random matrix theory has also been applied to jammed solids, which are widely studied nowadays~\cite{Beltukov-random-matrix-theory-2015, Altieri-jamming-glass-transitions-2019, Narayan-vibrational-spectrum-granular-2024}. Among the various random matrix ensembles, the Wishart ensemble is crucial for examining the characteristics of strongly disordered stable mechanical systems, as it involves positive-semidefinite random matrices. In particular, the correlated Wishart ensemble allowed us to derive the analytical form of the boson peak and the dynamical structure factor~\cite{Beltukov-iofferegel-criterion-diffusion-2013, Conyuh-random-matrix-approach-2021}, describe the Ioffe-Regel crossover and the viscoelastic properties~\cite{Conyuh-iofferegel-criterion-viscoelastic-2021}. It also explains the enhancement of the local elastic moduli due to the suppression of the nonaffine deformations in the interfacial region around nanoparticles in an amorphous host material, which may significantly increase the macroscopic stiffness of nanocomposites~\cite{Beltukov-local-elastic-properties-2022, Conyuh-effective-elastic-moduli-2023}. The obtained thickness of the interfacial region depends on the strength of the disorder and is determined by the heterogeneity length scale $ξ$, which is of the same order as the length scale of the boson peak. 

In this work, we present a theoretical study of nonaffine elastic response correlation properties using the theory of positive-definite correlated random matrices and compare them to the results of numerical simulations. Section~\ref{sec:NA} defines the main concepts and the averaging procedure for obtaining the covariance of nonaffine displacements. In Section~\ref{sec:RMT}, the random matrix theory is applied to find the general form of the covariance of nonaffine deformations. In Section~\ref{sec:StatProp}, the case of an amorphous medium with homogeneous statistical properties is considered. In Section~\ref{sec:SDM}, the case of strong disorder is considered, and the main results for the correlation properties of the divergence and the rotor of the nonaffine displacement field are obtained.
In Section~\ref{sec:num}, the theoretical results are compared to numerical models: the rigidity percolation model and molecular dynamics simulations of amorphous polystyrene and
the Lennard-Jones glass. 
Finally, the results obtained are discussed in Section~\ref{sec:Disc}.

\section{Affine and nonaffine displacements}
\label{sec:NA}
In this paper, we explore the mechanical behavior of an amorphous solid that has been quenched to zero temperature, allowing it to reach a local potential energy minimum. In the absence of thermal fluctuations, the system can persist in this state for an extended period. When external forces $\V f_i$ are applied at a frequency $ω$, they induce particle displacements $\V u_i$ from their equilibrium positions. We consider small external forces, ensuring that the quenched system stays near the local equilibrium without transitioning to other potential energy minima. In the linear approximation, the elastic response $\V u_i$ is defined by the following system of linear equations:
\begin{equation}
    \sum_{jβ} \left[Φ_{iα,jβ} - ω^2m_{iα,jβ} \right]u_{jβ} = f_{iα},
    \label{eq:lin-resp}
\end{equation}
where $\M{Φ}$ is the force constant matrix and $\M m$ is the mass matrix. The indices $i$ and $j$ enumerate the atoms in the system ($1\ldots N$), while $α$ and $β$ denote the Cartesian indices ($x,y,z$ for $d=3$ or $x,y$ for $d=2$) of the corresponding atoms. The elements of the force-constant matrix $\M{Φ}$ are defined by the Hessian of the total potential energy $U$
\begin{equation}
    Φ_{iα,jβ} = \frac{∂^2 U(\V r_1, \V r_2, \ldots, \V r_N)}{∂r_{iα}∂r_{jβ}}\bigg|_{\V r_i = \V r_i^{\rm eq}}
    \label{eq:Phi}
\end{equation}
taken at the equilibrium atomic coordinates $\V r_i^{\rm eq}$. The mass matrix $\M m$ is the Hessian of the kinetic energy with respect to velocities, specifically $m_{iα,jβ} = m_iδ_{ij}δ_{αβ}$ for the atomic system under consideration. Utilizing Eq.~(\ref{eq:lin-resp}), the atomic displacements can be explicitly expressed as
\begin{equation}
    u_{iα} = \sum_{jβ} \left(\frac{1}{\M{Φ} - ω^2\M m}\right)_{iα,jβ}f_{jβ}^{}.
    \label{eq:u}
\end{equation}
Matrices $\M{Φ}$ and $\M m$ are of size $N_{\rm dof} \times N_{\rm dof}$, where $N_{\rm dof} = d\thin N$ is the number of degrees of freedom. However, there are some number $N_{\rm triv}$ of trivial degrees of freedom related to the free translation and rotation of the system  without changing the potential energy ($d$ and $d(d-1)/2$, respectively).
In systems subject to periodic boundary conditions, global rotations are effectively suppressed, so such systems exhibit only $N_{\rm triv} = d$ trivial (translational) degrees of freedom. 
Therefore, to simplify the analysis, the inversion of any $N_{\rm dof}\times N_{\rm dof}$ matrix is assumed to be performed over the reduced subspace of nontrivial degrees of freedom $N_{\rm dof}' = N_{\rm dof} - N_{\rm triv}$. This helps to exclude trivial degrees of freedom and avoid divergence caused by the system's acceleration, which might formally arise due to nonzero total forces as $ω\to0$.

The specific equilibrium coordinates $\V r_1^{\rm eq}, \V r_2^{\rm eq}, \ldots, \V r_N^{\rm eq}$ depend on the cooling process of the melts forming the amorphous material. Therefore, the components of the force-constants matrix $\hat{Φ}$ depend on the particular system under consideration and can vary in a broad range~\cite{Alexander-amorphous-solids-their-1998}. 

As a result, under an external uniform stress, not only macroscopic (affine) deformations occur, but also local (nonaffine) atomic displacements that depend on the degree of disorder. In this light, it is natural to represent the atomic displacement field as a sum of affine and nonaffine components~\cite{DiDonna-nonaffine-correlations-random-2005}:
\begin{equation}
    u_{iα} = u^{\rm aff}_{iα} + u^{\rm naff}_{iα},
    \label{eq:u-naff}
\end{equation}
where the affine displacements are linearly defined by the macroscopic uniform strain tensor $\T{ε}$ as follows:
\begin{equation}
    u^{\rm aff}_{iα} = \sum_{β} ε_{αβ} r_{iβ}^{\rm eq}.
    \label{eq:strain-tensor}
\end{equation}
As nonaffine deformations arise due to the system's disorder, it is reasonable to assume that their average across the ensemble of the disordered system equals zero.

Each realization of the amorphous medium gives different atomic coordinates $\V r_i^{\rm eq}$. Therefore, the direct averaging of atomic displacements over an ensemble of such amorphous systems is not really meaningful.  To compare displacement fields across different realizations of the disordered solid, which have different atomic coordinates $\V r_i^{\rm eq}$, we project the atomic displacements onto a fixed reference lattice with nodes $\V r_i^{\rm ref}$ placed on a simple cubic lattice for $d=3$ or a square lattice for $d=2$. For the simplicity of the further analysis, we assume that the number of nodes is equal to the number of atoms $N$. Therefore, the average density of atoms $n_{\rm at}$ is equal to the density of the nodes. This allows us to define a displacement field $u_{jα}^{\rm ref}$ that is comparable across the ensemble:
\begin{equation}
    u_{jα}^{\rm ref} = \sum_i u_{iα} ϕ_{ij},   \label{eq:ref_map}
\end{equation}
where $\M{ϕ}$ is some smoothing matrix satisfying the normalization condition $\sum_i ϕ_{ij}=1$ and the rule $\sum_i ϕ_{ij}(\Vr_i - \Vr_j^{\rm ref})=0$ for all $j$. The latter guarantees that the displacements $u_{jα}^{\rm ref}$ of the reference lattice are affine if the atomic displacements $u_{jα}$ are affine. The further analysis weakly depends on the particular choice of $\M{ϕ}$. An example of such a smoothing matrix is provided in Section~\ref{sec:MDS} devoted to molecular dynamics. 

At the same time, forces $f_{iα}$ acting on atoms placed in some force field may also depend on their positions. Therefore, it is natural to define them using  
force density
defined on the nodes $f_{jα}^{\rm ref}$, which do not depend on the specific equilibrium coordinates:
\begin{equation}
    f_{iα} = \frac{1}{n_{\rm at}}\sum_j ϕ_{ij} f_{jα}^{\rm ref}, 
\end{equation}
where the same smoothing matrix $\M{ϕ}$ is used. Thus, we define the force-constant
density
matrix and the 
mass
density
matrix on the reference lattice as
\begin{align}
    Φ_{iα,jβ}^{\rm ref} &= n_{\rm at} \sum_{i'j'}\bigl(\M{ϕ}^{-1}\bigr)_{ii'}Φ_{i'α,j'β}^\p\bigl(\M{ϕ}^{-1}\bigr)_{j'j}, \\
    m_{iα,jβ}^{\rm ref} &= n_{\rm at} \sum_{i'j'}\bigl(\M{ϕ}^{-1}\bigr)_{ii'}m_{i'α,j'β}^\p\bigl(\M{ϕ}^{-1}\bigr)_{j'j}.
\end{align}
The formulation of forces and associated quantities per unit volume on a reference lattice makes the further transition to the continuum limit particularly clear.
Such definitions ensure the symmetry of the matrices $\M{Φ}^{\rm ref}$ and $\M m^{\rm ref}$ and fulfill the expected relationship
\begin{equation}
    u_{iα}^{\rm ref} = \sum_{jβ} \left(\frac{1}{\displaystyle \M{Φ}^{\rm ref} - ω^2\M m^{\rm ref}}\right)_{iα,jβ}f_{jβ}^{\rm ref}. 
    \label{eq:uf_ref}
\end{equation}

According to Eq.~\eqref{eq:uf_ref}, the average response $\avgs{u_{iα}^{\rm ref}}$ can be written as:
\begin{equation}
    \avg{u_{iα}^{\rm ref}} = \sum_{jβ} G_{iα,jβ} f_{jβ}^{\rm ref}, 
    \label{eq:u-aff}
\end{equation}
where the resolvent (also known as the Green function) is introduced:
\begin{equation}
    \MG = \avg{\frac{1}{\displaystyle \M{Φ}^{\rm ref} - ω^2 \M m^{\rm ref}}}.
\end{equation}
Angular brackets denote the averaging over different realizations of the amorphous system. The resolvent $\MG$ is a regular matrix on a simple cubic (or square) lattice. For the further analysis of static elastic properties, we will use small imaginary frequency $ω=iϵ$ with small positive $ϵ\to0$, which results in the positive-definite resolvent
\begin{equation}
    \MG = \avg{\frac{1}{\displaystyle \M{Φ}^{\rm ref} + ϵ^2 \M m^{\rm ref}}}.
    \label{eq:G}
\end{equation}
For uniform stress, the average displacements of the reference nodes $⟨u^{\rm ref}_{iα}⟩$ are affine:
\begin{equation}
    ⟨u^{\rm ref}_{iα}⟩ = \sum_{β} ε_{αβ} r_{iβ}^{\rm ref}.  \label{eq:affine-ref}
\end{equation}
Therefore, the nonaffine displacements are the deviation from the mean displacement, 
\begin{equation}
    u_α^{\rm naff}\big(\Vr_i^{\rm ref}\big) ≡ u^{\rm ref}_{iα} - ⟨u^{\rm ref}_{iα}⟩.
\end{equation}
Thus, nonaffine displacements have a zero mean but may possess nonzero correlations, which represent the correlation properties of the elastic response of an amorphous solid. The pairwise covariance of nonaffine displacements is defined as
\begin{equation}
    K_{iα,jβ} = \avg{u^{\rm ref}_{iα}\,u^{{\rm ref}}_{jβ}} - ⟨u_{iα}^{\rm ref}⟩ ⟨u_{jβ}^{\rm ref}⟩.
    \label{eq:naCorr-1}
\end{equation}
In this paper, we first analyze the covariance matrix $K_{iα,jβ}$ for arbitrary (presumably smooth) force fields given by $f^{\rm ref}_{iα}$. Then, in the final steps, we will take into account the property (\ref{eq:affine-ref}) for the affine strain.

To facilitate a more concise notation for subsequent analysis, we substitute the combined indices $iα$ and $jβ$ with the simpler indices $a$ and $b$, respectively. In this manner, the covariance of nonaffine displacements can be written as follows:
\begin{equation}
    K_{ab} = ⟨u_a^{\rm ref}u_b^{\rm ref}⟩ - ⟨u_a^{\rm ref}⟩⟨u_b^{\rm ref}⟩.
\end{equation}
The average product of displacements can be written as
\begin{equation}
    ⟨u_a^{\rm ref} u_b^{\rm ref}⟩ = \sum_{a'b'}\F G_{ab,a'b'} f_{a'}^{\rm ref} f_{b'}^{\rm ref},
\end{equation}
where the four-point resolvent (also known as the two-particle Green function) is introduced:
\begin{equation}
    \F G_{ab,a'b'} = \avg{
        \left(\frac{1}{\displaystyle \M{Φ}^{\rm ref} + ϵ^2 \M m^{\rm ref}}\right)_{\!aa'}
        \left(\frac{1}{\displaystyle \M{Φ}^{\rm ref} + ϵ^2 \M m^{\rm ref}}\right)_{\!bb'}
    }.  \label{eq:tpG}
\end{equation}

The primary objective of this paper is to examine the spatial properties of the covariance of nonaffine displacements given by $K_{ab}$. To facilitate the averaging and derive $\MG$ and $\FM G$, we employ techniques of the random matrix theory.

\section{Random matrix theory}
\label{sec:RMT}
\subsection{Correlated Wishart ensemble}

The essential aspect of employing random matrix theory in analyzing the mechanical properties of solids is ensuring their mechanical stability, which is represented by the force-constant matrix $\M{Φ}$ and consequently the matrix $\M{Φ}^{\rm ref}$ being positive semidefinite. This implies that the total potential energy 
\begin{equation}
    U = \frac{1}{2} \sum_{ab} Φ_{ab}^{\rm ref} u_a^{\rm ref} u_b^{\rm ref}
\end{equation}
remains nonnegative for any atomic displacements $u_a$. There is also a permutation rule $Φ_{ab} = Φ_{ba}$ which follows from the definition of the force constant matrix \eqref{eq:Phi}. The above-mentioned conditions are equivalent to the possibility of representing the force-constant matrix as \cite{Beltukov-iofferegel-criterion-diffusion-2013}:
\begin{equation}
    Φ_{ab}^{\rm ref} = \sum_{k} A_{ak} A_{bk},
    \label{eq:AAT}
\end{equation}
which can be written as $\M{Φ}^{\rm ref} = \M A · \M A^T$, where the dot ($·$) means the contraction over one inner index. The matrix $\M A$ in the relation \eqref{eq:AAT} can be interpreted as it follows. Its index $a$ enumerates degrees of freedom, while its index $k$ enumerates bonds with the positive-definite quadratic potential energy
\begin{equation}
    U_k = \frac{1}{2} \biggl(\sum_a A_{ak} u_a^{\rm ref}\biggr)^2. \label{eq:Uk}
\end{equation}
Each bond may involve several degrees of freedom, and the number and positions of nonzero elements in the matrix $\M A$ depend on the type of interaction between atoms in an amorphous solid \cite{Beltukov-boson-peak-various-2016}. Thus $\M A$ is a rectangular matrix of size $N_{\rm dof} \times N_{\rm b}$, where $N_{\rm dof} = d \thin N$ is the number of degrees of freedom, and $N_{\rm b}$ is the number of bonds.
For the further analysis, we reserve indices $a$ and $b$ to enumerate degrees of freedom and indices $k$ and $l$ to enumerate bonds. 
The structure of Eqs.~(\ref{eq:AAT}) and (\ref{eq:Uk}) is universal for mechanically stable systems. If certain bonds exhibit negative-definite (and thus unstable) quadratic potential energy, it is always possible to construct an alternative combination of bonds that is stable, see Appendix~\ref{app:pot} for further details.

The presence of disorder in atomic arrangements leads to the random nature of the matrix $\M A$. Therefore, to describe the amorphous state, one can assume that the nonzero matrix elements $A_{ak}$ are random numbers. Following the paper \cite{Conyuh-effective-elastic-moduli-2023}, we consider correlated Gaussian random numbers $A_{ak}$ with zero mean and covariance 
\begin{equation}
    \avg{A_{ak} A_{bl}} = \C_{ab,kl}.
    \label{eq:C-corr}
\end{equation}
For such a random matrix $\M A$, ensemble of matrices $\M{Φ}^{\rm ref} = \M A · \M A^T$ forms a correlated Wishart ensemble.

The covariance matrix $\FM C$ as well as the two-particle Green function $\FM G$ have four distinct indices. These matrices are referred to as four-point matrices, in accordance with the notation employed in~\cite{Eichmann-bound-states-spectral-2024}. In our paper, we use calligraphic letters and wide hats for four-point matrices to distinguish them from standard two-index matrices.
The covariance matrix $\FM C$ plays a key role in describing the general properties of strongly disordered systems since it contains their main features. 
In real amorphous solids, short-range interactions between atoms dominate over long-range interactions, which results in a sparse force-constant matrix $\M{Φ}$. Therefore, the covariance matrix $\FM C$ also has a short-range sparse structure. The random matrix theory remains applicable to such random sparse matrices as long as each particle interacts with $z$ other particles and $zd\gg 1$, as discussed in Appendix~\ref{sec:aver}.

The covariance matrix $\FM C$ has several important properties, which follow from the definition~(\ref{eq:C-corr}). In particular,
\begin{equation}
    \sum_{abkl} \C_{ab,kl} B_{ak} B_{bl} = \avg{\Tr^2\bigl[\M A · \M B^T\bigr]} \ge 0
\end{equation}
for any real matrix $\M B$ of size $N_{\rm dof} \times N_{\rm b}$. It follows that for any positive-semidefinite matrix $\M X$ of size $N_{\rm b} \times N_{\rm b}$, the matrix $\M Y = \FM C : \M X$ of size $N_{\rm dof} \times N_{\rm dof}$ is also positive-semidefinite, whereas the double contraction ($:$) means the summation over two inner indices
\begin{equation}
    Y_{ab} = \sum_{kl} \C_{ab,kl} X_{kl}.
\end{equation}
Indeed, any positive-semidefinite matrix $\M X$ can be presented as $\M X = \M Z·\M Z^T$, and
\begin{equation}
    \sum_{ab} x_a Y_{ab} x_b = \sum_m \sum_{abkl} x_a Z_{km} \C_{ab,kl} Z_{lm} x_b \ge 0
\end{equation}
for any real $x_a$. Thus $\M X \mapsto \FM C:\M X$ is a linear map from matrices of size $N_{\rm b}\times N_{\rm b}$ to matrices of size $N_{\rm dof}\times N_{\rm dof}$, which preserves positive semidefiniteness. Using the same idea, one can show that for any positive-semidefinite matrix $\M Y$, the matrix $\M X = \FM C^{\,T}:\M Y$ is also positive semidefinite, where the transposition of the four-point matrix is the interchange of the left and the right pair of indices. Additionally, for any given pair of positive-semidefinite matrices, denoted $\M Y$ and $\M X$, the expression $\M Y : \FM C : \M X$ evaluates to a number that is guaranteed to be nonnegative.

As demonstrated in \cite{Conyuh-effective-elastic-moduli-2023} using the diagrammatic technique, the covariance matrix $\FM C$ determines the resolvent $\MG$ defined in Eq.~(\ref{eq:G}) by the coupled Dyson-Schwinger equations:
\begin{align}
    \M G &= \frac{1}{\FM C:\MGs + ϵ^2\M m^{\rm ref}}, 
    \label{eq:G1}
    \\
    \MGs &= \frac{1}{\FM C^T\!:\MG + \M1}, 
    \label{eq:G-star}
\end{align}
where $\MGs$ represents the secondary resolvent of the size $N_{\rm b}\times N_{\rm b}$:
\begin{equation}
    \MGs = \avg{\frac{1}{\M A^T \frac{1}{ϵ^2\M m^{\rm ref}} \M A +  \M1}}.
\end{equation}
In Eqs. (\ref{eq:G1}) and (\ref{eq:G-star}) we have neglected the fluctuations in the mass matrix, $\M m^{\rm ref} = ⟨\M m^{\rm ref}⟩$. While some amorphous systems may have significant mass disorder, the actual mass distribution is not relevant in the static limit $ϵ\to0$. Therefore, the subsequent results are not affected by the assumption about mass disorder.

Equations (\ref{eq:G1}) and (\ref{eq:G-star}) form a system of nonlinear equations. This naturally raises a question regarding whether a solution to the system indeed exists and, if it does, whether such a solution is unique. The physical solution corresponds to the positive-semidefinite resolvent $\M G$ for $ϵ > 0$. Since the covariance matrix $\FM C$ as well as the matrix inversion preserves positive semidefiniteness, the matrix $\MGs$ is also positive semidefinite.

Using the block-diagonal resolvent $\MG_{\rm bd} = \bigl(\begin{smallmatrix}\MG &0\\0&\MGs\end{smallmatrix}\bigr)$, one can write Eqs.~(\ref{eq:G1}) and (\ref{eq:G-star}) as one Dyson-Schwinger equation
\begin{equation}
    \MG_{\rm bd} = \frac{1}{\FM C_{\rm bd}:\MG_{\rm bd} + \M V_{\rm bd}},  \label{eq:Dyson2}
\end{equation}
where
\begin{equation}
    \FM C_{\rm bd}:\Bigl(\begin{smallmatrix}\MG & 0\\0&\MGs\end{smallmatrix}\Bigr) = \Bigl(\begin{smallmatrix}\FM C:\MGs &0\\0&\FM C^{\,T}\!:\MG\end{smallmatrix}\Bigr), \quad 
    \M V_{\rm bd} = \Bigl(\begin{smallmatrix}ϵ^2\M m &0\\0&\M1\end{smallmatrix}\Bigr).
\end{equation}
The four-point matrix $\FM C_{\rm bd}$ preserves positive-semidefiniteness, and the matrices $\MG_{\rm bd}$ and $\M V_{\rm bd}$ are positive-definite. It is known that there is one and only one solution of Eq.~(\ref{eq:Dyson2}) in the domain of positive-definite matrices \cite{Helton-operatorvalued-semicircular-elements-2007}. In order to obtain the solution, one can use the simple iteration procedure. In terms of the resolvent $\MG$, this iteration writes
\begin{equation}
    \MG_{n+1} = \Bigl(\FM C:\bigl(\FM C^{\,T}\!:\MG_n + \M1\bigr)^{-1} + ϵ^2\M m^{\rm ref}\Bigr)^{-1}.   \label{eq:Dyson_iter}
\end{equation}
It converges to the solution, $\lim_{n\to\infty}\MG_n = \MG$, for any starting resolvent $\MG_0$ being a positive-semidefinite matrix \cite{Helton-operatorvalued-semicircular-elements-2007}.

\subsection{Four-point resolvent}

Employing the diagram method outlined in Appendix~\ref{sec:aver}, the four-point resolvent $\FM G$ can be expressed as
\begin{equation}
    \F G_{ab,a'b'} = \F L_{ab,a'b'} + \F L_{ab',a'b} - \F R_{ab,a'b'},  \label{eq:tpG2}
\end{equation}
where the four-point matrix $\FM L$ is defined as the solution of the equation (known as the two-particle Dyson equation)
\begin{equation}
    \FM L = \FM R + \FM R:\FM T:\FM L,\label{eq:tpDyson}
\end{equation}
where
\begin{align}
    \F R_{ab,a'b'} &= G_{aa'}G_{bb'},  \label{eq:R_def} \\
    \F T_{ab,a'b'} &= \sum_{klk'l'}\C_{ab,kl}\Gs_{kk'}\Gs_{ll'}\C_{a'b',k'l'}.  \label{eq:T_def}
\end{align}
The first term in Eq.~(\ref{eq:tpG2}) corresponds to the ladder diagrams, while the second term corresponds to the twisted diagrams (see Appendix~\ref{sec:aver}). Accordingly, the covariance of the nonaffine displacements can be written as
\begin{equation}
    K_{ab} = K_{ab}^{\rm ld} + K_{ab}^{\rm tw},
\end{equation}
where the ladder term is
\begin{equation}
    K_{ab}^{\rm ld} = \sum_{a'b'}(\F L_{ab,a'b'} - \F R_{ab,a'b'})f_{a'}f_{b'},
\end{equation}
and twisted term is
\begin{equation}
    K_{ab}^{\rm tw} = \sum_{a'b'}(\F L_{ab',a'b} - \F R_{ab,a'b'})f_{a'}f_{b'}.
\end{equation}

\subsection{Four-point eigenvalue decomposition}
\label{subsec:eigval}

Four-point matrices $\FM R$ and $\FM T$ are symmetric with respect to the transposition of the left and right pair of indices, $\FM R=\FM R^T$, $\FM T=\FM T^T$. At the same time, they are semidefinite with respect to the double contraction:
\begin{equation}
    \M X:\FM R:\M X \ge 0, \quad \M X:\FM T:\M X \ge 0
\end{equation}
for any matrix $\M X$, which follows from the definitions (\ref{eq:R_def}) and (\ref{eq:T_def}). Such positive-semidefinite self-adjoint linear operators acting on matrices
(also known as superoperators)
can be examined through the eigenvalue analysis. One can find $N_{\rm dof}^2$ eigenvalues $θ_v \ge 0$ such that
\begin{equation}
    \FM T:\FM R:\M S^{(v)} = θ_v \M S^{(v)},  \label{eq:eigval_TR}
\end{equation}
where $\M S^{(v)}$ are the basis matrices (also known as the eigenmatrices \cite{Ajanki-stability-matrix-dyson-2019}), which are orthonormal with the weight $\FM R$ in the following sense 
\begin{equation}
    \M S^{(v)} : \FM R : \M S^{(v')\dag} \equiv \Tr\bigl[\M S^{(v)}·\MG·\M S^{(v')\dag}·\MG\bigr]= N'_{\rm dof}δ_{vv'}.   \label{eq:orthog}
\end{equation}
The Hermitian conjugation ($\dag$) is employed in order to use the complex basis $\M S^{(v)}$. Although real four-point matrices $\FM T$ and $\FM R$ allow for a real basis, the complex basis is chosen to enable further application of Fourier analysis.

At the same time, one can write the four-point matrices $\FM T$ and $\FM R$ as
\begin{align}
    \F T_{ab,a'b'} &= \frac{1}{N'_{\rm dof}}\sum_v S_{ab}^{(v)} θ_v S_{a'b'}^{(v)\dag}\,,  \label{eq:eigval_T} \\
    \F R_{ab,a'b'} &= \frac{1}{N'_{\rm dof}}\sum_v \bigl(\MG·\M S^{(v)}·\MG\bigr)_{ab} \bigl(\MG·\M S^{(v)\dag}·\MG\bigr)_{a'b'}\,.  \label{eq:eigval_R}
\end{align}
Consequently, the ladder four-point matrix $\FM L$ is derived through the eigenvalue decomposition (\ref{eq:eigval_T}):
\begin{equation}
    \F L_{ab,a'b'} = \frac{1}{N'_{\rm dof}}\sum_v \bigl(\MG·\M S^{(v)}·\MG\bigr)_{ab}\,\frac{1}{1 - θ_v}\,\bigl(\MG·\M S^{(v)\dag}·\MG\bigr)_{a'b'}\,.
\end{equation}
Thus, the nonaffine displacement covariance $K_{ab}$ has two components:
\begin{align}
    K^{\rm ld}_{ab} &= \frac{1}{N'_{\rm dof}}\sum_{v} \bigl(\MG·\M S^{(v)}·\MG\bigr)_{ab}\,\frac{θ_v }{1 - θ_v}\,\bigl(\overline{u}·\M S^{(v)\dag}·\overline{u}\bigr)\,, 
    \label{eq:ld_basis}\\
    K^{\rm tw}_{ab} &= \frac{1}{N'_{\rm dof}}\sum_{v} \bigl(\MG·\M S^{(v)}·\overline{u}\bigr)_a\,\frac{θ_v}{1 - θ_v}\,\bigl(\overline{u}·\M S^{(v)\dag}·\MG\bigr)_b\,,
    \label{eq:tw_basis}
\end{align}
where  $\overline{u}$ denotes  a vector composed of $N_{\rm dof}$ elements $⟨u_a^{\rm ref}⟩$.
There is an important difference between the two components: the terms in the twisted component $K^{\rm tw}_{ab}$ have the form of a direct product of two $N_{\rm dof}$-dimensional vectors indexed by $a$ and $b$, while the terms in the ladder component $K^{\rm ld}_{ab}$ cannot be decomposed in such a way.

One can expect that $θ_v$ cannot exceed 1, which means the ladder four-point matrix $\FM L$ is positive semidefinite. To elaborate on this question, one can consider the final convergence of the iteration (\ref{eq:Dyson_iter}). In this case, $\MG_n = \MG + \dd\MG_n$ with small $\dd\MG_n$. The next step of the iteration is $\MG_{n+1} = \MG + \dd\MG_{n+1}$ with
\begin{equation}
    \dd\MG_{n+1} = \FM R:\FM T:\dd\MG_n.  \label{eq:iter_dG}
\end{equation}
The iteration converges, which means that all eigenvalues of $\FM R:\FM T$ and $\FM T:\FM R$ are less than 1 for any $ϵ > 0$. Therefore, $0\le θ_v < 1$.
At the same time, some of the limiting values $\lim_{ϵ\to0}θ_v$ may be equal to 1.

For the further analysis, we assume that the correlation matrix $\FM C$ is such that for any positive-definite matrices $\M X$ and $\M Y$, the matrices $\FM C : \M X$ and $\FM C^{\,T} : \M Y$ are strictly positive definite, excluding $N_{\rm triv}$ trivial degrees of freedom. This assumption is essential and posits that the disorder introduced by the correlation matrix $\FM C$ influences all $N_{\rm dof}'$ nontrivial degrees of freedom and all $N_{\rm b}$ bonds (at the same time, the matrix $\FM C$ may be highly sparse). In this case, the map $\M X \mapsto \FM T : \FM R : \M X$ preserves the positive-definiteness of the matrix $\M X$ in the subspace of nontrivial degrees of freedom. For such maps, which preserve the cone of positive-definiteness, the Perron–Frobenius theorem is applicable \cite{Ajanki-stability-matrix-dyson-2019}. It implies that the largest eigenvalue, denoted by $θ_0$, is nondegenerate and the corresponding basis matrix $\M S^{(0)}$ can be chosen to be positive semidefinite, having $N_{\rm dof}'$ positive eigenvalues and $N_{\rm triv}$ zero eigenvalues. 

Taking the inverse of the Dyson-Schwinger equations (\ref{eq:G1}), (\ref{eq:G-star}), and multiplying them by $\MG$ and $\MGs$ with the double contraction, we obtain
\begin{align}
    \MG:\FM C:\MGs + ϵ^2 (\MG:\M m) &= N_{\rm dof}', \\
    \MGs:\FM C^{\,T}:\MG + \Tr[\MGs] &= N_{\rm b},
\end{align}
where we have taken into account that $\MG:\MG^{-1} = N_{\rm dof}'$ due to the exclusion of $N_{\rm triv}$ trivial degrees of freedom from the inversion. The difference of the above equations gives
\begin{equation}
    \Tr[\MGs] = N_{\rm b} - N_{\rm dof}' + ϵ^2 (\MG:\M m).
\end{equation}
If $N_{\rm b} > N_{\rm dof}'$, $\Tr[\MGs] = N_{\rm b} - N'_{\rm dof}$ as $ϵ\to 0$. So the mean of eigenvalues of $\MGs$ is equal to 
\begin{equation}
    ϰ = 1 - N_{\rm dof}'/N_{\rm b}.  \label{eq:kappa}
\end{equation}
The parameter $ϰ$ plays a crucial role in the proposed theory. The case $ϰ = 0$ corresponds to the minimum number of bonds required in a stiff system, given by the Maxwell counting rule~\cite{Maxwell-calculation-equilibrium-stiffness-1864}, also known as the isostatic condition. It quantifies the distance from isostaticity and serves as an inverse measure of the disorder strength, with $ϰ→0$ corresponding to the critical, strongly disordered limit.

The case of $N_{\rm b} < N_{\rm dof}'$ and $ϰ<0$ is drastically different. The matrices $\MG$, $\MGs$, and $\M m$ are positive semidefinite. Consequently, $\Tr[\MGs]$ and $(\MG:\M m)$ cannot be negative. As such, the value of $(\MG:\M m)$ diverges proportionally to $ϵ^{-2}$. Hence, according to the Dyson-Schwinger equations (\ref{eq:G1})--(\ref{eq:G-star}), the resolvents scale as $\MG \sim ϵ^{-2}$ and $\MGs \sim ϵ^2$. Such resolvents correspond to a very loose system with an insufficient number of bonds, which we will not consider further.

Of greater interest is the case of a critical, isostatic system when $N_{\rm b} = N_{\rm dof}'$ and $ϰ=0$. In this case, $\Tr[\MGs] = ϵ^2 (\MG:\M m)$ and the resolvents scale as $\MG \sim ϵ^{-1}$ and $\MGs\sim ϵ$ according to the Dyson-Schwinger equations (\ref{eq:G1})--(\ref{eq:G-star}). For such a scaling, the Dyson-Schwinger equation reduces to $\MG = \bigl(\FM C:\bigl(\FM C^{\,T}:\MG\bigr){}^{-1}\bigr){}^{-1}$. Due to the ability of this equation to multiply $\MG$ by any scaling factor, there exists the basis matrix $\M S^{(0)}$, which approaches $\MG^{-1}$ as $ϵ\to0$, and the corresponding eigenvalue $θ_0$, which approaches 1. Using $\MG^{-1}$ as the first approximation of the basis matrix $\M S^{(0)}$, the eigenvalue $θ_0$ can be estimated using Eq.~(\ref{eq:eigval_TR}) and the orthogonality relation (\ref{eq:orthog}) as
\begin{equation}
    θ_0 \approx \frac{\MG^{-1}:\FM R:\FM T:\FM R:\MG^{-1}}{\MG^{-1}:\FM R:\MG^{-1}}\,.  \label{eq:theta_approx}
\end{equation}
Using the properties $\FM R:\MG^{-1} = \MG$, $\FM C^{\,T}:\MG = (\MGs)^{-1} - \M1$, and the definition of the matrix $\FM T$ given by Eq.~(\ref{eq:T_def}), the eigenvalue is
\begin{equation}
    θ_0 \approx \frac{\Tr\bigl[(\M1 - \MGs)^2\bigr]}{N_{\rm dof}'} \approx 1 - 2 \frac{\Tr\bigl[\MGs\bigr]}{N_{\rm dof}'}.
\end{equation}
Therefore, the eigenvalue $θ_0$ is close to 1 and $1 - θ_0 \sim ϵ$. 

Behavior of the near-critical system with $0 < ϰ\ll 1$ 
and $ϵ → 0$
is similar to that of the critical system with finite $ϵ$
and $ϰ=0$.
If the correlation properties of the disorder are sufficiently homogeneous, the maximum eigenvalue of $\MGs$ is of the order of the mean eigenvalue $ϰ$. 
Using $\MG^{-1}$ as the first approximation of the basis matrix $\M S^{(0)}$ and Eq.~(\ref{eq:theta_approx}), we obtain
\begin{equation}
    θ_0 \approx \frac{\Tr\bigl[(\M1 - \MGs)^2\bigr]}{N_{\rm dof}'} \approx \frac{N_{\rm b}}{N_{\rm dof}'} - 2 \frac{\Tr\bigl[\MGs\bigr]}{N_{\rm dof}'} = \frac{1-2ϰ}{1-ϰ}.
\end{equation}
Therefore, the eigenvalue $θ_0$ is close to 1 and $1 - θ_0 \sim ϰ$. 
Consequently, one can expect that all scaling properties for the critical system remain valid when $ϵ$ is replaced by~$ϰ$:
\begin{equation}
    \bigl\| \MG \bigr\| \sim ϰ^{-1}, \quad \bigl\| \MGs \bigr\| \sim ϰ, \quad 1 - θ_0 \sim ϰ. \label{eq:scale_G_kappa}
\end{equation}
Such scalings are based on the assumption that the disorder is evenly distributed over the degrees of freedom. In general, if some degrees of freedom have stronger disorder than others, the approximation $\M S^{(0)} \approx \MG^{-1}$ may be inaccurate. Our further analysis does not strictly rely on this approximation, but it can give an estimate for the properties of $\M S^{(0)}$.

In real amorphous materials, the variety of atomic interactions leads to a wide distribution of bond stiffness. Consequently, a simple Maxwell counting of bonds is insufficient; one must account for the efficiency of these bonds in constraining the system’s degrees of freedom. However, the calculation of the number of effective bonds is indeed non-trivial and may significantly differ for low-coordinated (like linear polymers) and high-coordinated (like metallic glasses) systems. Thus, the parameter $ϰ$ should be viewed further as an effective inverse measure of disorder strength that accounts for the distribution of bond stiffness and the efficiency of atomic constraints in the system. As will be seen further, the objective measure of disorder is given by the largest eigenvalue as $\tk = 1 - θ_0$.

\section{Amorphous medium with homogeneous statistical properties}
\label{sec:StatProp}
In the previous Section, the general properties of the correlated Wishart ensemble are discussed. However, the properties of the mechanical system impose restrictions on the matrix $\M A$. In this Section, the indices of degrees of freedom $a$ and $b$ are replaced back by the combined indices $iα$ and $jβ$, indicating that one atom has $d$ degrees of freedom.

Since the total energy of the system $U$ does not change when the system is translated or rotated as a whole, there are the sum rules for the matrix $\M A$. As it follows from the Eqs.~\eqref{eq:Phi}, \eqref{eq:AAT}, the translational invariance when the whole system is shifted to a constant vector $u_{iα} = u_α$ leads to the sum rule 
\begin{equation}
    \sum_i A_{iα,k} = 0
    \label{eq:Arule-trans}
\end{equation}
for any Cartesian index $α$ and the bond number $k$. When the system is rotated by an infinitesimal angle, the displacements of the atoms are expressed as $u_{iα} = \sum_β ω_{αβ} r_{iβ}^{\rm ref}$, where $ω_{αβ}$ are elements of the infinitesimal rotation tensor \cite{Maradudin-theory-lattice-dynamics-1971}. Therefore, as it follows from the Eqs.~\eqref{eq:Phi} and \eqref{eq:AAT}, the rotational invariance leads to the rule
\begin{equation}
    \sum_i \LC_{αβγ}^{\vphantom{\rm ref}} r_{iβ}^{\rm ref}A_{iα,k}^{\vphantom{\rm ref}}  = 0
    \label{eq:Arule-rot}
\end{equation}
for any bond number $k$ and Cartesian index $γ$, where $\LC_{αβγ}$ is the Levi-Civita symbol.

The obtained relations for the matrix $\M A$ impose corresponding restrictions on the form of the correlation matrix $\FM C$:
\begin{gather}
    \sum_i \F C_{iαjβ,kl} = \sum_j \F C_{iαjβ,kl} = 0, 
    \label{eq:C_sumrule_transl}
    \\
    \sum_i \LC_{αγη}^{\vphantom{\rm ref}} r_{iγ}^{\rm ref}\F C_{iαjβ,kl}^{\vphantom{\rm ref}}  = \sum_j \LC_{βγη}^{\vphantom{\rm ref}} r_{jγ}^{\rm ref}\F C_{iαjβ,kl}^{\vphantom{\rm ref}}  = 0. 
    \label{eq:C_sumrule_rot}
\end{gather}
As a result, these sum rules are also applied for the four-point matrix $\F T_{iαjβ,i'α'j'β'}$ with the summation taken over any of the four indices $i$, $j$, $i'$, $j'$.  All basis matrices also obey the given sum rules: 
\begin{gather}
    \sum_i S^{(v)}_{iαjβ} = \sum_j S^{(v)}_{iαjβ} = 0,  \label{eq:sumrule_S}\\
    \sum_i \LC_{αγη}r_{iγ}^{\rm ref}S^{(v)}_{iαjβ} = \sum_j \LC_{βγη}r_{jγ}^{\rm ref}S^{(v)}_{iαjβ} = 0.
    \label{eq:rotsumrule_S}
\end{gather}

The amorphous medium under consideration has homogeneous statistical properties. As a result, the resolvent $G_{iαjβ}$ depends only on the difference between coordinates $\V r_i - \V r_j$. Thus, it is worth defining the Green function as a Fourier transform of the resolvent
\begin{equation}
    G_{αβ}(\V q) = \frac{1}{N}\sum_{ij} G_{iαjβ} \, e^{i\V q·(\V r_i^{\rm ref} - \V r_j^{\rm ref})}.
\end{equation}
Similarly, the Fourier transform of mean atomic displacements and forces is:
\begin{align}
    ⟨u_α^{\rm ref}⟩(\V q) &= \sum_i ⟨u_{iα}^{\rm ref}⟩ e^{i\V q·\V r_i^{\rm ref}},
    \\
    f_α^{\rm ref}(\V q) &= \sum_i f_{iα}^{\rm ref} e^{i\V q·\V r_i^{\rm ref}}.
\end{align}
Therefore, Eq.~(\ref{eq:u-aff}) reads
\begin{equation}
    ⟨u_α^{\rm ref}⟩(\V q) = G_{αβ}(\V q) f_β^{\rm ref}(\V q),
\end{equation}
where the Einstein notation for Cartesian indices is employed. 
In the continuum limit ($q\ll 1/a_0$, where $a_0$ is the reference lattice constant), the Green function of an isotropic elastic medium is~\cite{Mura-micromechanics-defects-solids-1987}
\begin{equation}
    G_{αβ}(\V q) = \frac{1}{λ+2μ}\frac{q_αq_β}{q^4} + \frac{1}{\mu}\left(\frac{δ_{αβ}}{q^2} - \frac{q_αq_β}{q^4}\right), \label{eq:Green} 
\end{equation}
where $λ$ and $μ$ are Lam\'e moduli of the medium. They are related to other known elastic moduli such that $μ$ is the shear modulus, $λ + \frac{2}{3}μ$ is the bulk modulus, and $λ/(2λ+2μ)$ is the Poisson's ratio for $d=3$. For  $d=2$, the relation is slightly different: $μ$ is the shear modulus, $λ + μ$ is the bulk modulus, and $λ/(λ+2μ)$ is the Poisson's ratio.

The backward relation between $f_α^{\rm ref}(\V q)$ and $⟨u_β^{\rm ref}⟩(\V q)$ is given by the inverse Green function, which is given by the tensor inversion of Eq.~(\ref{eq:Green}):
\begin{equation}
    (G^{-1})_{αβ}(\V q) = (λ+μ) q_α q_β + μ\thin δ_{αβ}q^2.   \label{eq:invGreen} 
\end{equation}
For zero frequency, $ϵ\to 0$, the inverse Green function $(G^{-1})_{αβ}(\V q)$ is nothing more than the Fourier transform of the effective force-constant matrix $\avgs{(\M{Φ}^{\rm ref})^{-1}}{}^{-1}$.

Due to the statistical homogeneity of the system, the shift $\V r_i^{\rm ref} \to \V r_i^{\rm ref} + \V{δr}$ leaves the system unchanged, where $\V{δr}$ is the vector connecting two arbitrary nodes of the reference lattice. Such a shift may only multiply the basis matrix $\M S^{(v)}$ by some phase factor $e^{-i\V p·\V{δr}}$, where the wavevector $\V p$ depends on the eigenvalue index $v$. This property is analogous to Bloch's theorem \cite{Ashcroft-solid-state-physics-1976, Esposito-exactly-solvable-model-2005}. Therefore, all the eigenvalues can be grouped by the wavevector $\V p$, and there is a one-to-one correspondence between the index $v$ and the pair $(n\V p)$, where $n$ is the branch (band) number. 
Using this notation, we can write the resulting property of the basis matrices as
\begin{equation}
    \sum_{\smash{ij}} S^{(n\Vp)}_{iαjβ}\,e^{i\Vq·(\Vr_i^{\rm ref} - \Vr_j^{\rm ref}) + i\Vp'·(\Vr_i^{\rm ref} + \Vr_j^{\rm ref})/2} = 0
    \label{eq:selection_p}
\end{equation}
if $\Vp'≠\Vp$.

\begin{figure}
    \centering
    \includegraphics[scale=0.7]{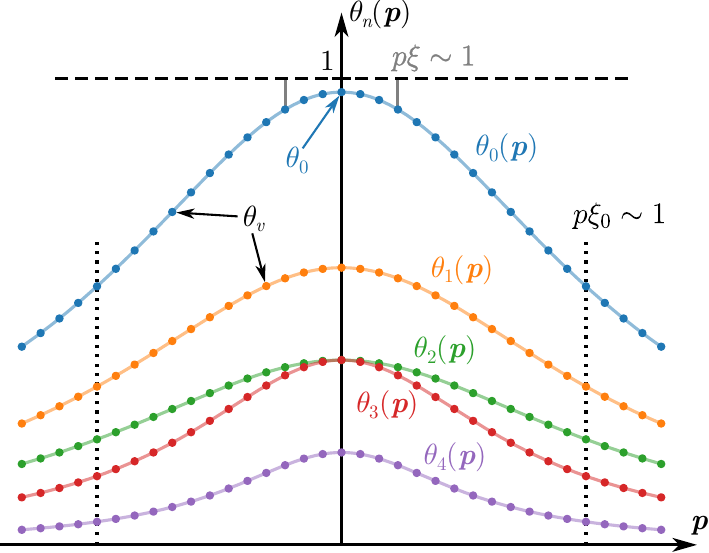}
    \caption{Schematic illustration of the  branches  $θ_n(\Vp)$ (color lines) and discrete eigenvalues $θ_v$ for the finite system (color points) for $\tk = 0.03$. Vertical gray lines show the position of $pξ\sim 1$. The vertical dotted lines show the position of $pξ_0\sim 1$.}
    \label{fig:branches}
\end{figure}

The branches $θ_{n}(\V p) \equiv θ_{(n\V p)}$ are illustrated in Fig.~\ref{fig:branches}. According to the Perron–Frobenius theorem, the largest eigenvalue $θ_0$ is not degenerate. Therefore, it may correspond to $\Vp=0$ only and $θ_0(0) = θ_0$. Otherwise, there are at least two identical eigenvalues, which correspond to $\Vp$ and $-\Vp$. The Perron–Frobenius theorem is applicable also for the subspace of the operator $\FM T$ given by $\Vp = 0$. Therefore, the upper branch $θ_0(\Vp)$ is not degenerate for $\Vp=0$ and all other branches $θ_n(\Vp)$ are lower than $θ_0(\Vp)$ for $\Vp=0$ and $n>0$. At the same time, some of the branches $θ_n(\Vp)$ may be degenerate for $\Vp=0$ and $n > 0$, as illustrated in Fig.~\ref{fig:branches}.

To analyze nonaffine deformations, we introduce the Fourier transform of basis matrices
\begin{equation}
    S^{(n)}_{αβ}(\V q_1, \V q_2) = \frac{1}{N}\sum_{ij\Vp} S^{(n\Vp)}_{iαjβ}\,e^{i\V q_1·\V r_i^{\rm ref} - i\Vq_2·\V r_j^{\rm ref}},
    \label{eq:S_Fourier}
\end{equation}
which encodes the complete structure of the $n$-th branch in reciprocal space. The wavevector $\Vp = \Vq_1 - \Vq_2$ is chosen automatically by Eq.~(\ref{eq:selection_p}). Within this reciprocal-space representation, the translational and rotational sum rules
given by Eqs.~(\ref{eq:sumrule_S}) and (\ref{eq:rotsumrule_S}) take the form of the following identities
\begin{gather}
    S^{(n)}_{iαjβ}(0, \Vq_2) = S^{(n)}_{iαjβ}(\Vq_1, 0) = 0,
    \label{eq:sumrule_S_q1_q2}
    \\[1mm]
    \LC_{αγη}^{\vphantom{\rm ref}}\frac{∂S_{αβ}^{(n)}(\Vq_1, \Vq_2)}{∂ q_{1γ}}\bigg|_{\Vq_1 = 0} = 
    \LC_{βγη}\frac{∂S_{αβ}^{(n)}(\Vq_1, \Vq_2)}{∂ q_{2γ}}\bigg|_{\Vq_2 = 0} = 0.
    \label{eq:rotsumrule_S_q1_q2}
\end{gather}
\vspace{-2mm}

Given that $⟨u_{iα}^{\rm ref}⟩$ represents the affine displacements $u_{iα}^{\rm aff}$ as defined in Eq.~(\ref{eq:strain-tensor}), the covariance of nonaffine deformation in reciprocal space is expressed as
\begin{multline}
    K_{αβ}(\V q) ≡ \int \big\langle u^{\rm naff}_α(\V r)\,u^{\rm naff}_β(0)\big\rangle\,e^{i\Vq·\Vr} \dd\Vr
    \\
    = \frac{1}{N}\sum_{ij} K_{iαjβ}\,e^{i\V q·(\V r_i^{\rm ref} - \V r_j^{\rm ref})},
    \label{s:fourier_K}
\end{multline}
which can be decomposed into two distinct contributions:
\begin{equation}
    K_{αβ}(\V q) = K^{\rm ld}_{αβ}(\V q) + K^{\rm tw}_{αβ}(\V q)
\end{equation}
corresponding to the ladder and twisted terms discussed in the previous Section. As shown in Appendix~\ref{app:fourier}, both contributions have the form
\begin{equation}
    K^{\rm ld/tw}_{αβ}(\V q) = G_{αα'}(\V q)H^{\rm ld/tw}_{α'β'}(\Vq)G_{β'β}(\V q).
    \label{eq:K_ld_tw_q}
\end{equation}
For the ladder contribution, we obtain
\begin{equation}
    H^{\rm ld}_{αβ}(\V q) = \sum_{n} h_n^\p S^{(n)}_{αβ}(\V q, \Vq),
\end{equation}
where
\begin{equation}
    h_n = \frac{θ_n(0)}{1 - θ_n(0)} \frac{ε_{γγ'}^\p ε_{ηη'}^\p}{d} \frac{∂^2 S^{(n)*}_{γη}(\Vq_1, \Vq_2)}{∂q_{1γ'} ∂q_{2η'}}\biggr|_{\substack{\V q_1 = 0\\\Vq_2=0}}.
    \label{eq:hn}
\end{equation}
For the twisted contribution, we find
\begin{equation}
    H^{\rm tw}_{αβ}(\V q) = \frac{1}{d}\sum_{n}F_{α}^{(n)}(\V q) \frac{θ_n(\V q) }{1 - θ_n(\V q)} F_{β}^{(n)*}(\V q),   \label{eq:tw_space}
\end{equation}
where
\begin{equation}
    F_{α}^{(n)}(\V q) = ε_{γγ'}^\p\frac{∂S^{(n)}_{αγ}(\Vq, \Vq_2)}{∂q_{2γ'}}\biggr|_{\Vq_2 = 0}.
    \label{eq:Fα_q}
\end{equation}
The properties of $K^{\rm ld}_{αβ}(\V q)$ and $K^{\rm tw}_{αβ}(\V q)$ inherit the properties given by Eqs.~(\ref{eq:ld_basis}) and (\ref{eq:tw_basis}): the terms in $K^{\rm tw}_{αβ}(\V q)$ are direct products of two Cartesian vectors, while the terms in $K^{\rm ld}_{αβ}(\V q)$ cannot be decomposed in such a way.

\section{Strongly disordered medium}
\label{sec:SDM}
As was discussed at the end of Section~\ref{subsec:eigval}, a strongly disordered amorphous medium corresponds to the case $ϰ \ll 1$. In this case, $θ_0(0)=θ_0$ is close to 1. All other branches $θ_n(0)$ are lower than $θ_0(0)$. Therefore, we can assume that in the general case, only the upper branch is close to 1 (see Fig.~\ref{fig:branches}). Near $\V p = 0$, it can be approximated as
\begin{equation}
    θ_0(\V p) = 1 - \tk - \xi_0^2 p^2  \label{eq:theta_series}
\end{equation}
with $\tk\sim ϰ\ll1$. In the general case, the length scale $\xi_0$ does not contain any critical behavior and has a value of the order of the length scale of the atomic interaction, which is usually about the interatomic scale. The approximation (\ref{eq:theta_series}) is applicable while $\xi_0p\ll 1$, otherwise more terms are required.

For small $\Vq_1$ and $\Vq_2$, the tensor $S^\spar{0}_{αβ}(\Vq_1, \Vq_2)$ is isotropic. Due to the translational and rotational identities (\ref{eq:sumrule_S_q1_q2}) and (\ref{eq:rotsumrule_S_q1_q2}) it has the following form (see Appendix~\ref{app:eigenmatrices} for more details):
\begin{equation}
    S^{(0)}_{αβ}(\V q_1, \V q_2) = \tl\,q_{1α}^\p q_{2β}^\p + \tm\, q_{1β}^\p q_{2α}^\p + \tm\, δ_{\mkern-0.5mu αβ}^\p q_{1γ}^\p q_{2γ}^\p,
    \label{eq:S0_ab_q_p}
\end{equation}
where the parameters $\tl$ and $\tm$ can be viewed as the disorder Lam\'e moduli. If the approximation $\M S^\spar{0} \approx \MG^{-1}$ is valid (see Section \ref{subsec:eigval}), then $S^\spar{0}_{αβ}(\V q, \V q) \approx (G^{-1})_{αβ}(\V q)$ 
and the disorder moduli coincide with the elastic moduli: $\tm\approx μ$ and \mbox{$\tl \approx λ$}. As an example of such a case, in Appendix~\ref{sec:Uncorr} a simple model of uncorrelated disorder is considered. However, we prefer to consider a more general case here.

As a result, the ladder term of the covariances is
\begin{equation}
    K^{\rm ld}_{αβ}(\V q) = \frac{h_0(\tl+2\tm)}{(λ+2μ)^2}\frac{q_αq_β}{q^4} + \frac{h_0\tm}{μ^2}\left( \frac{δ_{αβ}}{q^2} - \frac{q_αq_β}{q^4}\right),
    \label{eq:Kladder(q)}
\end{equation}
where
\begin{equation}
    h_0 = \frac{\tl(ε_{γγ})^2 + 2\tm ε_{γη}ε_{γη}}{d\thin \tk}.
\end{equation}
It is worth noting that the numerator in this expression has the same structure as the elastic energy density of an isotropic solid subject to a strain field $ε_{αβ}$.

For the twisted term, we have
\begin{equation}
    K^{\rm tw}_{αβ}(\V q) = \frac{G_{αα'}(\V q)F_{α'}^{(0)}(\V q)F_{β'}^{(0)*}(\V q) G_{β'β}(\V q)}{d\thin\tk(1 + ξ^2q^2)},
    \label{eq:Ktwisted(q)}
\end{equation}
where
\begin{gather}
    ξ = ξ_0/\sqrt{\tk},  \label{eq:xi} \\
    F_{α}^{(0)}(\V q) = \tl q_αε_{γγ} + 2\tm ε_{αγ} q_γ.
\end{gather}
As $ϰ$ and $\tk$ approach zero, the length scale $\xi$ becomes indefinitely large, potentially surpassing all other length scales within the system, including the correlation length of the disorder.
The twisted term $K^{\rm tw}_{αβ}(\V q)$ given by Eq.~\ref{eq:Ktwisted(q)} is a direct product of two vectors, which follows from the structure of Eqs.~(\ref{eq:tw_basis}) and (\ref{eq:tw_space}). As we will see in the following subsections, it will imply special symmetry properties of this term.

The spatial correlations of the nonaffine displacements can be computed using the inverse Fourier transform
\begin{multline}
    \quad K_{αβ}(\V r) ≡ \big\langle u^{\rm naff}_α(\V r)\,u^{\rm naff}_β(0)\big\rangle 
    \\
    = \frac{1}{V_b}\int_{V_b} K_{αβ}(\V q)e^{-i\V q·\V r} \dd\Vq \quad 
    \label{eq:Fourier}
\end{multline}
with the integral taken over the first Brillouin zone of the reference lattice, where the volume of this zone is given by $V_b = n_{\rm at}(2π)^d$. Since $K_{αβ}(\Vq) = K_{αβ}^{\rm lad}(\Vq) + K_{αβ}^{\rm tw}(\Vq)$, the inverse Fourier transform (\ref{eq:Fourier}) can be performed for both terms independently. 
In the following analysis, we focus on the long-range regime characterized by $q a_0 \ll 1$ and $r \gg a_0$. Under these conditions, the detailed microscopic structure of the reference lattice does not play a significant role and can therefore be neglected.
For the ladder term given by Eq.~(\ref{eq:Kladder(q)}) we obtain:
\begin{align}
    K_{αβ}^{\rm ld}(\V r) &= -b_+δ_{αβ}\ln\frac{r}{r_0} + b_-\frac{r_αr_β}{r^2}, \quad d=2,\label{eq:Kld_2d}\\
    K_{αβ}^{\rm ld}(\V r) &= b_+\frac{δ_{αβ}}{2r} + b_-\frac{r_αr_β}{2r^3}, \quad d=3, \label{eq:Kld_3d}
\end{align}
where $r_0$ is the normalization length, which depends on the system size, and
\begin{equation}
    b_{\pm} = \frac{h_0}{4πn_{\rm at}} \left( \frac{\tm}{μ^2}\pm\frac{\tl+2\tm}{(λ+2μ)^2} \right).
\end{equation}
This power-law decay $K_{αβ}^{\rm ld}(\V r) \propto r^{2-d}$ is in agreement with the main results of the work \cite{DiDonna-nonaffine-correlations-random-2005} and has the same spatial behavior as the Green function of an isotropic elastic medium~\cite{Mura-micromechanics-defects-solids-1987}. However, the power-law decay does not contain any specific length scale.

In contrast, the twisted term $K_{αβ}^{\rm tw}(\Vq)$ given by Eq.~(\ref{eq:Ktwisted(q)}) has a nontrivial length scale $\xi$ due to $1 + ξ^2q^2$ in the denominator. The corresponding spatial correlation function $K_{αβ}^{\rm tw}$ has exponential terms ${\sim}e^{-r/ξ}$ along with the power-law terms, while the expression is much lengthier and presented in Appendix~\ref{app:cor}.

Thus, spatial correlations of nonaffine displacements $K_{αβ}(\V r) = ⟨u^{\rm naff}_α(\V r)\,u^{\rm naff}_β(0)⟩$ contain both power-law and exponential decay terms even in the case of uncorrelated disorder. In previous theoretical studies, to the best of our knowledge, the exponential decay was obtained for the correlated disorder only \cite{DiDonna-nonaffine-correlations-random-2005}.

In reality, the derivatives of displacements usually play a more important role than the displacements themselves, since they represent the local strain. Therefore, we calculate the correlation function of the divergence of the nonaffine displacement field, along with the correlation function of the rotor. Remarkably, our analysis shows that these functions have a simple analytical form.

\subsection{Divergence}

As a result of deformation, correlations between variations in the density of matter can occur in an amorphous system. This corresponds to the divergence correlator of nonaffine displacements of the form
\begin{equation}
    K_{\rm div}(\V r - \V r') = \avg{\div \V u^{\rm naff}(\Vr)\, \div \V u^{\rm naff}(\Vr')},
    \label{eq:Kdiv-1}
\end{equation}
where the divergence of nonaffine displacements on the reference lattice is defined as the corresponding finite difference. It is easy to see that the divergence correlator \eqref{eq:Kdiv-1} is expressed by the correlator \eqref{eq:Fourier} in the following form:
\begin{equation}
    K_{\rm div}(\V r-\V r') = \frac{∂^2 K_{α β}(\Vr-\Vr')}{∂r_{α} ∂r'_{β}}.
    \label{eq:Kdiv-2}
\end{equation}
In the reciprocal space, it corresponds to
\begin{equation}
    K_{\rm div}(\V q) = q_{α} q_{β} K_{α β}(\V q).
    \label{eq:Kdiv-q1}
\end{equation}
For the volumetric deformation $ε_{αβ} = εδ_{αβ}$, we obtain
\begin{equation}
    K_{\rm div}(\V q) = c_1 + \frac{c_2}{1+ξ^2q^2}, 
    \label{eq:Kdiv-q}
\end{equation}
where $c_1$ and $c_2$ are numerical factors of the order of~$ε^2/\tk$:%
\begin{align}
    c_1 &= \frac{(\tl + 2\tm)(d\tl + 2\tm)}{(λ+2μ)^2\tk} ε^2,\\
    c_2 &= \frac{(d\tl + 2\tm)^2}{d(λ+2μ)^2\tk}ε^2,
\end{align}
where $\tl + 2\tm/d$ has the meaning of the disorder bulk modulus in any dimensions. 
Taking the Fourier transform, we obtain
\begin{equation}
    K_{\rm div}(\V r) = \frac{c_1}{n_{\rm at}}δ(\V r) + \frac{c_2}{n_{\rm at}}D(r)
    \label{eq:Kdiv-r}
\end{equation}
with
\begin{align}
    D(r) &= \frac{{\rm K}_0(r/ξ)}{2\pi ξ^2}, \quad d=2, \\
    D(r) &= \frac{e^{-r/ξ}}{4\pi rξ^2}, \quad d=3,  \label{eq:D_3d}
\end{align}
where ${\rm K}_0(x)$ is the modified Bessel function of the second kind. In three dimensions, $D(r)$ has the explicit exponential decay given by Eq.~(\ref{eq:D_3d}), while in two dimensions, it has the asymptotic exponential behavior $D(r) = e^{-r/\xi}/\sqrt{8πrξ^3}$ for $r \gg ξ$. 

One can see that the classical power-law decay of the correlator $K_{αβ}(\Vr)$ given by Eqs.~(\ref{eq:Kld_2d})--(\ref{eq:Kld_3d}) transforms to the delta function of the correlator of divergence $K_{\rm div}(\V r)$ in Eq.~(\ref{eq:Kdiv-r}). At the same time, the second term in Eq.~(\ref{eq:Kdiv-r}) has the exponential decay at the length scale $ξ$, which follows from Eq.~(\ref{eq:Ktwisted(q)}). Both terms have comparable integral contributions to $K_{\rm div}(\V r)$ given by
\begin{equation}
    \int K_{\rm div}(\Vr) \dd\Vr  = \frac{c_1 + c_2}{n_{\rm at}} \sim ξ^2.
\end{equation}
This integral is proportional to $ξ^2$ for any dimension $d$, which is in agreement with the previous works~\cite{Lerner-breakdown-continuum-elasticity-2014, Yan-variational-arguments-vibrational-2016, Shimada-spatial-structure-quasilocalized-2018, Lerner-anomalous-linear-elasticity-2023}. 

For an arbitrary strain tensor $\T{ε}$, the correlation function is anisotropic and has a lengthy expression given in Appendix~\ref{app:cor}. However, the isotropic part of $K_{\rm div}(\Vr)$ has the same form as Eq.~(\ref{eq:Kdiv-r}):
\begin{equation}
    K_{\rm div}^{\rm iso}(\V r) = \frac{c_1}{n_{\rm at}}δ(\V r) + \frac{c_2}{n_{\rm at}}D(r),
    \label{eq:Kdiv-r-iso}
\end{equation}
where the coefficients $c_1$ and $c_2$ for the general strain tensor $\T{ε}$ are presented in Eqs.~(\ref{eq:c1_gen}) and (\ref{eq:c2_iso}).

\subsection{Rotor}

One can also calculate the correlation function of the rotor of the nonaffine displacement field
\begin{equation}
    K_{\rm rot}(\V r-\V r') = \avg{\rot \V u^{\rm naff}(\Vr) · \rot \V u^{\rm naff}(\Vr')},
    \label{eq:Krot}
\end{equation}
which gives
\begin{equation}
    K_{\rm rot}(\V q) = (q^2 δ_{αβ} - q_{α} q_{β}) K_{α β}(\V q).
    \label{eq:Krot-q1}
\end{equation}
For the volumetric deformation $ε_{αβ} = εδ_{αβ}$, we obtain
\begin{equation}
    K_{\rm rot}(\V q) = c_3,
\end{equation}
where $c_3$ is a numerical factor of the order of~$ε^2/\tk$:
\begin{equation}
    c_3 = \frac{\tm(d\tl +2\tm)}{μ^2\tk}(d-1)ε^2. 
\end{equation}
As a result,
\begin{equation}
    K_{\rm rot}(\V r) = \frac{c_3}{n_{\rm at}}δ(\V r).
    \label{eq:Krot-r}
\end{equation}
It is notable that the correlator of rotors does not contain the exponential decay in the case of volumetric deformation. 
This conclusion arises from the observation that the twisted contribution $K^{\rm tw}_{αβ}(\V q)$ can be expressed as a direct product of two vectors, together with the fact that the upper branch $θ_0(\V p)$ under study is nondegenerate and exhibits isotropic symmetry at $\Vp=0$. Consequently, for an isotropic (volumetric) strain, one obtains the structure $K^{\rm tw}_{αβ}(\V q) = q_α q_β f(q)$, which does not contribute to the rotor correlation function $K_{\rm rot}(\V q)$ defined in Eq.~(\ref{eq:Krot-q1}).

In practice, the delta function in Eq.~(\ref{eq:Krot-r}) is not precise and exhibits broadening on the atomic scale. Furthermore, our analysis is limited to the upper branch $θ_0(\V p)$. Nonetheless, certain lower branches might contribute a nonzero exponential term in $K_{\rm rot}(\Vr)$. These branches are not anticipated to exhibit critical behavior and similarly possess an atomic length scale. As a result, the proposed theory predicts that the correlation length for the rotor field is shorter than that for the divergence field, which was not expected in advance. 
This outcome will be verified in Section~\ref{sec:num} using numerical examples. Furthermore, certain lower branches may give rise to a small power-law tail, as will be discussed in the next subsection.

For an arbitrary strain tensor $\T{ε}$, the correlation function $K_{\rm rot}(\Vq)$ has additional terms, which depend on $\Vq$, see Appendix~\ref{app:cor}. Therefore, the isotropic part of $K_{\rm rot}(\Vr)$ has both the delta function and the additional exponential term in the general case:
\begin{equation}
    K_{\rm rot}^{\rm iso}(\V r) = \frac{c_3}{n_{\rm at}}δ(\V r) + \frac{c_4}{n_{\rm at}}D(r),
    \label{eq:Krot-r-iso}
\end{equation}
where the coefficients $c_3$ and $c_4$ for the general strain tensor $\T{ε}$ are presented in Eqs.~(\ref{eq:c3_gen}) and (\ref{eq:c4_iso}). The coefficient $c_4$ is equal to zero for the volumetric strain only.

\subsection{Influence of lower branches}
The upper branch $θ_0(\Vp)$ is the closest to 1 (see Fig.~\ref{fig:branches}), which provides the largest contribution to correlation functions that is proportional to $1/(1-θ_0(\Vp))$ and diverges as $1/\tk$. The contribution from lower branches $θ_n(\Vp)$ for $n > 0$ is generally less important. However, there is a small but nonzero contribution that is absent for the upper branch.

According to the Perron-Frobenius theorem, the upper branch $θ_0(\Vp)$ is non-degenerate for $\Vp = 0$. However, some lower branches may be degenerate, as illustrated in Fig.~\ref{fig:branches} by $θ_2(0)$ and $θ_3(0)$, and they have lower symmetry. The eigenvalues $θ_n(0)$ are analogous to the eigenenergies (with the minus sign) of a hydrogen atom: the ground state is a nondegenerate (apart from spin degeneracy) symmetric $s$-orbital, whereas the other states include degenerate $p$-, $d$-, and higher orbitals with greater energies. Depending on the parity of the orbital angular momentum quantum number, the corresponding orbitals exhibit either even (symmetric) or odd (antisymmetric) behavior under spatial inversion.

While this analogy is useful for developing qualitative intuition, a more rigorous symmetry-based classification of the basis matrices is provided in Appendix~\ref{app:eigenmatrices}. There it is shown that there are even and odd branches, and odd branches admit a linear-type expansion for small wavevectors $\Vq_1$ and $\Vq_2$:
\begin{equation}
    S^{(n)}_{αβ}(\Vq_1, \Vq_2) = V_{αβγ}^{(n)}(\Vp)(q_{1γ}^\p + q_{2γ}^\p),
    \label{eq:S_odd}
\end{equation}
where $\Vp = \Vq_1 - \Vq_2$. As demonstrated in Appendix~\ref{app:eigenmatrices}, the tensor $V_{αβγ}^\spar{n}(\Vp)$ is completely symmetric with respect to its three indices, is orthogonal to the vector $\Vp$,
\begin{equation}
    V_{αβγ}^{(n)}(\Vp)\,p_γ = 0,
    \label{eq:V_orthog}
\end{equation}
and, for sufficiently small values of $|\Vp|$, depends solely on the direction of the wavevector $\Vp$. Additionally, the odd branches correspond to antisymmetric basis matrices, and these matrices do not contribute to the ladder term given by Eq.~(\ref{eq:ld_basis}). However, there is a nontrivial contribution to the twisted term of the nonaffine correlation function $K_{αβ}^{\rm tw}(\Vq)$.

Using Eqs.~(\ref{eq:Fα_q}) and (\ref{eq:S_odd_der2}) for the small wavevector $\Vq$, we obtain
\begin{equation}
    F_α^{(n)}(\V q) = 2 V_{αγη}^{(n)}(\V q) ε_{γη},
\end{equation}
which gives
\begin{equation}
    H_{αβ}^{\rm tw}(\V q) = \frac{4}{d}{\sum_{n}}' V_{αγη}^{(n)}(\V q) ε_{γη}\frac{θ_n(0)}{1 - θ_n(0)}V_{βγ'η'}^{(n)*}(\V q) ε_{γ'η'},
    \label{eq:Hab_V}
\end{equation}
where the summation is performed over odd branches only. In the above formula, we neglect the dependence of eigenvalues on the wavevector $\Vq$. The corresponding contributions to the correlation functions are 
\begin{align}
    K_{\rm div}^{\rm odd}(\Vq) &= \frac{q_α q_βH_{αβ}^{\rm tw}(\V q)}{(λ+2μ)^2q^4},
    \\
    K_{\rm rot}^{\rm odd}(\Vq) &= \frac{(q^2 δ_{αβ} - q_α q_β)H_{αβ}^{\rm tw}(\V q)}{μ^2q^4}.
\end{align}
Due to the orthogonality relation (\ref{eq:V_orthog}), the divergence correlation function $K_{\rm div}^{\rm odd}(\Vq)$ is zero. At the same time, the rotor correlation function has a nonnegative contribution. For an isotropic medium, its most general form can be expressed as
\begin{equation}
    K_{\rm rot}^{\rm odd}(\Vq) = \frac{\big(ν_1^{\vphantom{\perp}} - \frac{ν_2}{d-1}\big)^{\vphantom{\perp}}\big(ε_{αα}^\perp(\Vq)\big)^2 + ν_2^{\vphantom{\perp}} ε_{αβ}^{\perp}(\Vq)ε_{αβ}^{\perp}(\Vq)}{μ^2q^2},
\end{equation}
where $ν_1$ and $ν_2$ are two nonnegative constants, and $ε_{αβ}^\perp(\Vq)$ denotes the transverse component of the strain tensor
\begin{equation}
    ε_{αβ}^\perp(\Vq) = \left(δ_{αα'} - \frac{q_αq_{α'}}{q^2}\right)\left(δ_{ββ'} - \frac{q_βq_{β'}}{q^2}\right)ε_{α'β'}.
\end{equation}
In two spatial dimensions, there exists only a single direction orthogonal to a given wavevector $\Vq$, which implies that the odd-branch rotor correlation function is characterized by a single coefficient $ν_1$.

The isotropic part of the obtained $K_{\rm rot}^{\rm odd}(\Vq)$ is
\begin{equation}
    K_{\rm rot}^{\rm odd, iso}(\Vq) = \frac{\tilde{ν}_1(ε_{αα})^2 +\tilde{ν}_2ε_{αβ}ε_{αβ}}{μ^2q^2},
\end{equation}
where
\begin{align}
    \tilde{ν}_1 &= \frac{d^2-3}{d(d+2)}ν_1 - \frac{(d-2)(d+1)}{d(d+2)(d-1)}ν_2,
    \\
    \tilde{ν}_2 &= \frac{2}{d(d+2)}ν_1 + \frac{(d-2)(d+1)}{(d+2)(d-1)}ν_2.
\end{align}
It results in the following power-law isotropic parts of the correlation functions in real space in three dimensions
\begin{equation}
    K_{\rm rot}^{\rm iso, odd}(\Vr) = \frac{(3ν_1-ν_2)(ε_{αα})^2 +(ν_1 + 3ν_2)ε_{αβ}ε_{αβ}}{30π n_{\rm at}μ^2}\frac{1}{r},
    \label{eq:K_rot_power}
\end{equation}
and logarithmic dependence in two dimensions
\begin{equation}
    K_{\rm rot}^{\rm iso, odd}(\Vr) = \frac{ν_1(ε_{αα})^2 +2ν_1ε_{αβ}ε_{αβ}}{16π n_{\rm at}μ^2}\ln \frac{r}{r_0}.
\end{equation}

The constants $ν_1$ and $ν_2$ depend on the eigenvalues $θ_n(0)$ of odd branches, which require $n>0$. These eigenvalues have lower values than $θ_0(0)$, and $1/(1 - θ_n(0))$ is not a large factor in Eq.~(\ref{eq:Hab_V}). A correlation between different bonds is required to have a nonzero $ν_1$ and $ν_2$ since $θ_n(0) = 0$ for the uncorrelated bonds (there is only an upper branch in this case, see Appendix~\ref{sec:Uncorr}). Using the analogy of eigenstates of the hydrogen atom, the eigenvalues $θ_n(0)$ are like the ionization energies, which are much smaller for the nonsymmetric orbitals than for the symmetric ground state.

Thus, nonzero $ν_1$ and $ν_2$ give rise to a small power-law contribution to the correlation functions $ K_{\rm div}^{\rm iso}(\Vr)$ and $ K_{\rm rot}^{\rm iso}(\Vr)$. This small contribution can be important at very large distances $r \gg ξ$, where the exponential term becomes negligible. These large-distance correlations can be observed in some numerical examples, as will be seen in the next Section.

\section{Numerical examples}
\label{sec:num}

\begin{figure}[t!]
    \centering
    \includegraphics[scale=0.8]{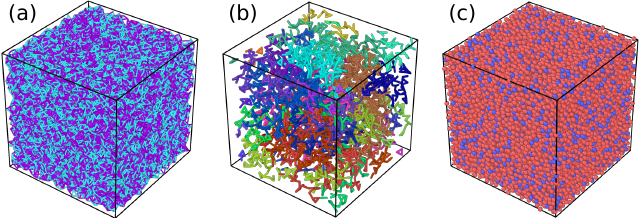}
    \caption{
    Numerical examples to investigate nonaffine displacements. (a) Rigidity percolation. Light blue lines represent cut bonds. (b) Molecular dynamics configuration of a model polystyrene in the amorphous state at zero temperature. Different colors represent different polymer chains. (c) Molecular dynamics configuration of a Lennard-Jones glass at zero temperature. Particles of types A and B are shown by red and blue colors, respectively. In each example, only 1/3 of the simulation cell in each direction is shown.
    }
    \label{fig:Systems}
\end{figure}

In order to validate the theoretical findings, numerical examples are considered, which include numerical investigations of a system near the rigidity percolation, molecular dynamics simulations of a model atactic polystyrene in the amorphous state, 
and molecular dynamics simulations of a Lennard-Jones glass, see Fig.~\ref{fig:Systems}(a-c).

\subsection{Rigidity percolation}
\label{sec:percolation}

Rigidity percolation is a generalization of the classical percolation problem to the case where the rigidity of an elastic network is studied instead of simple connectivity~\cite{Sahimi-vector-percolation-rigidity-2023, Thorpe-rigidity-percolation-1985, Jacobs-generic-rigidity-percolation-1995}. In such a case, each site (atom) has a vector quantity (displacement vector) instead of a scalar quantity. Thus, it is also known as a vector percolation problem. This concept applies to describe materials like gels~\cite{Zhang-correlated-rigidity-percolation-2019}, glasses~\cite{Thorpe-rigidity-percolation-glassy-1985}, and biological tissues~\cite{Petridou-rigidity-percolation-uncovers-2021}.

We consider a face-centered cubic (fcc) lattice with nearest neighbor atoms connected by harmonic springs. This lattice has 4 atoms in the unit cell, and each atom is connected to 12 neighboring atoms. The initial regular lattice with the neighboring distance $a_0$ is distorted: each atom is moved by the random vector $\V{δ}_i^{\rm rnd}$, whose components are uniformly distributed between $-a_0\sqrt{2}/4$ and $a_0\sqrt{2}/4$. It is the maximum displacement, which prevents the collision of neighboring atoms. The initial distortion of the lattice prevents atoms from being perfectly aligned along crystalline lines and resolves the associated issues, which are absent in actual amorphous solids~\cite{Jacobs-generic-rigidity-percolation-1995}.

\begin{figure}[t!]
    \centering
    \includegraphics[scale=0.8]{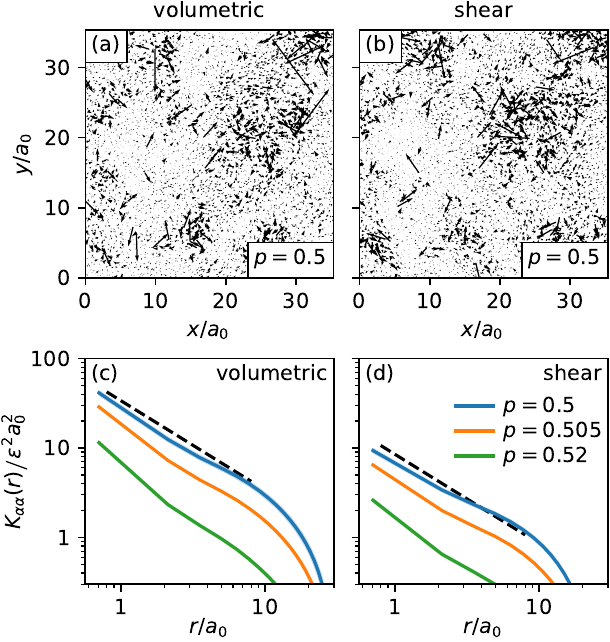}
    \caption{
    (a, b) Examples of nonaffine displacement fields in the rigidity percolation model under volumetric (a) and shear (b) strain. A slice of thickness $3 a_0$ in the $z$ direction is shown. In the directions $x$ and $y$ only half of the system is shown. 
    (c, d) Correlation function of the nonaffine displacement field $K_{αα}(r)$ in the rigidity percolation model system under volumetric (c) and shear (d) strain. 
    Dashed lines represent the dependence $1/r$ as a visual guide.
    }
    \label{fig:percolation_dr_dr}
\end{figure}

\begin{figure*}[t]
    \centering
    \includegraphics[scale=0.8]{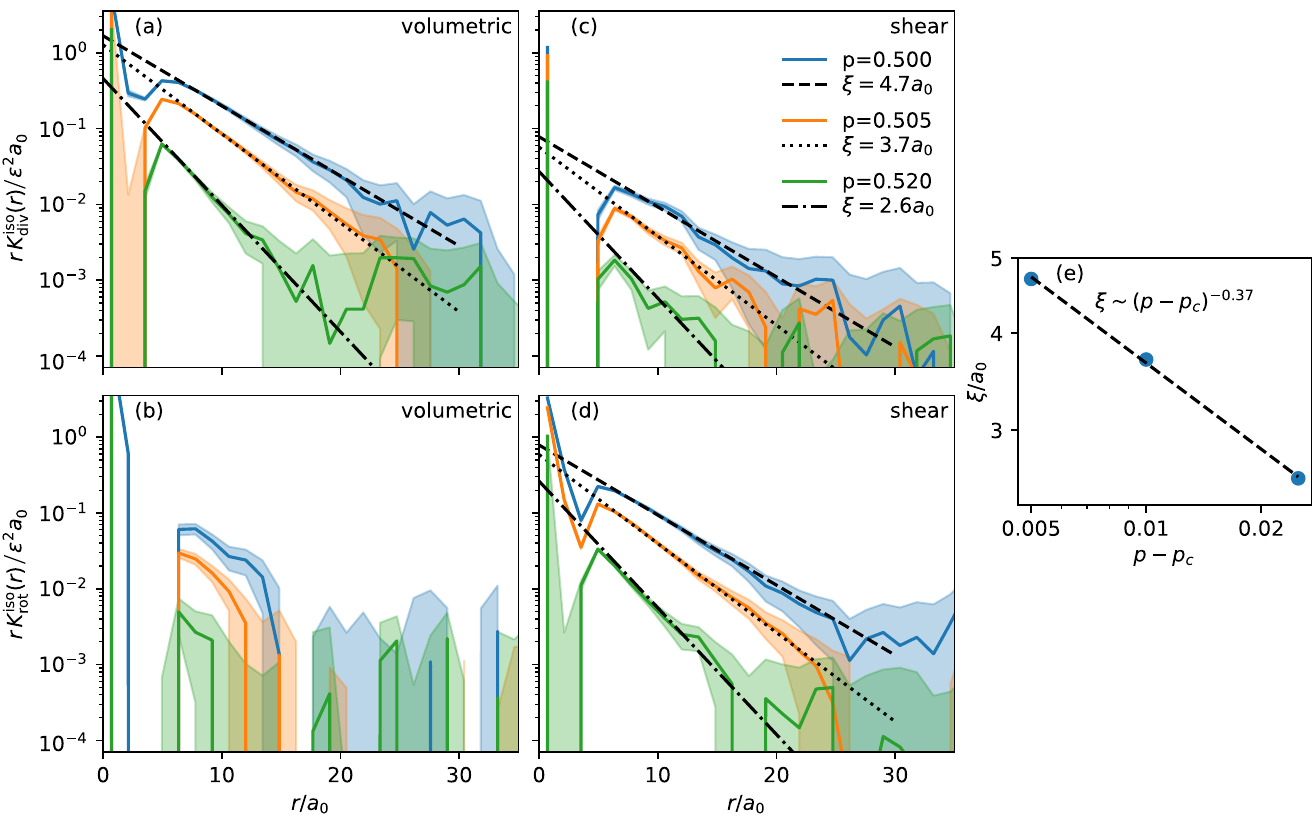}
    \caption{
    Correlation function of the divergence (a, c) and the rotor (b, d) of the nonaffine displacement field in the rigidity percolation model under volumetric (a, b) and shear (c, d) strain. The normalization factor $r/(ε^2a_0)$ is used to plot the data  and the isotropic part of the correlation function is shown. Shaded areas represent two standard deviations of the obtained data. Dashed and dotted lines represent the dependency $\exp(-r/ξ)$. (e) Dependence of the heterogeneity length scale $ξ$ on the vicinity of the percolation threshold $p - p_c$.
    }
    \label{fig:percolation_K_vs}
\end{figure*}

However, the main source of the disorder originates from the percolation: the neighboring atoms are connected with the probability $p$. Specifically, the pairwise potential energy of neighboring atoms in the distorted fcc lattice is
\begin{equation}
    U(r_{ij}) = \frac{k_{ij}}{2}\bigl(r_{ij} - r_{ij}^0\bigr)^2,
\end{equation}
where $r_{ij}^0$ is the distance between atoms $i$ and $j$ in the distorted lattice before any additional deformation is applied. The rigidity $k_{ij}$ is equal to $k_1$ with the probability $p$ and equal to near-zero value $k_0$ with the probability $1-p$. The nonzero rigidity of\, ``cut'' bonds $k_0 = 10^{-6}k_1$ is introduced to make the problem nonsingular. 

The lattice under consideration with $N$ atoms has $N_{\rm dof} = 3N$ degrees of freedom and $N_{\rm b} = 6pN$ bonds (bonds with near-zero rigidity $k_0$ are excluded from the counting). Thus, $p=0.5$ is a predicted value of the loss of rigidity. The actual value of the rigidity percolation threshold for the fcc lattice $p_c = 0.495$ is very close to the predicted value~\cite{Chubynsky-algorithms-threedimensional-rigidity-2007}. The probability $p$ is related to the previously defined parameter $ϰ$ in Eq.~(\ref{eq:kappa}) as $ϰ = 1 - p_c/p$. Therefore, close to the percolation threshold $ϰ ∼ p - p_c$, and the length scale diverges as $ξ \sim 1/\sqrt{ϰ} ∼ (p - p_c)^{-ν_{\rm na}}$ with $ν_{\rm na}=0.5$ according to the scaling given by Eq.~(\ref{eq:xi}). The presence of a large length scale $ξ$ near the percolation threshold and its dependence on $p - p_c$ will be studied numerically.

The distorted fcc lattice with $50^3$ unit cells ($N = 5\times10^5$ atoms) and periodic boundary conditions is considered for numerical evaluation slightly above the percolation threshold $p_c$. The systems with $p=0.500$, $p=0.505$, and $p=0.520$ were generated approximately 300 times. Two types of macroscopic deformations are applied to each of the systems: volumetric deformation and shear. For volumetric strain, the strain tensor is diagonal with diagonal elements $ε_{xx}=ε_{yy}=ε_{zz}=ε$. The shear deformation is represented by the traceless strain tensor $\T{ε}$. We selected it to be diagonal with diagonal elements $ε_{xx}= ε$ and $ε_{yy}=ε_{zz} = -ε/2$, allowing the use of an orthogonal periodic cell. For each infinitesimal strain $ε$, new equilibrium atomic positions are found by solving the linear system of equations.

Examples of the obtained nonaffine displacement fields close to the percolation threshold are shown in Fig.~\ref{fig:percolation_dr_dr}(a,b). Since the infinitesimal displacements have been calculated, the lengths of the arrows are not to scale. For visual clarity, the displacements of atoms with a coordination number 3 and smaller are not shown, since these atoms do not influence the rigidity of the system and some of them have enormous displacements.

The correlation function $K_{αα}(\V r) = ⟨u^{\rm naff}_α(\V r)\,u^{\rm naff}_α(0)⟩$ is plotted in Fig.~\ref{fig:percolation_dr_dr}(c,d) both for the volumetric strain and the shear. As was shown in the previous Section, the correlation function $K_{αα}(\V r)$ contains both exponential and power-law terms. For $r\gg ξ$, one can expect the long-range power-law tail $1/r$ until $r$ reaches the system size, which is observed in Fig.~\ref{fig:percolation_dr_dr}. As the parameter $p$ decreases, the magnitude of the 
correlation function 
$K_{αα}(\V r)$ increases, but no length scale is observed.

The most interesting information is contained in the correlation functions for the divergence and the rotor of the nonaffine deformations, $K_{\rm div}(\Vr)$ and $K_{\rm rot}(\Vr)$. Therefore, the divergence and the rotor of the nonaffine displacement field $\V u_j^{\rm ref}$ were computed using the finite-difference method. Then the corresponding correlation functions $K_{\rm div}^{\rm iso}(r)$ and $K_{\rm rot}^{\rm iso}(r)$ were obtained and averaged over different directions of $\Vr$ and all the generated systems.

The results are presented in Fig.~\ref{fig:percolation_K_vs} for $K_{\rm div}(\Vr)$ and $K_{\rm rot}(\Vr)$ for volumetric and shear deformations for different values of $p$. Equations (\ref{eq:Kdiv-r-iso}) and (\ref{eq:Krot-r-iso}) state that $K_{\rm div}^{\rm iso}(r)$ and $K_{\rm rot}^{\rm iso}(r)$ decay as $r^{-1}e^{-r/ξ}$ for large distances $r$. Therefore, we plot the normalized dimensionless correlation functions $r K_{\rm div/rot}^{\rm iso}(r)/ε^2a_0$ in Fig.~\ref{fig:percolation_K_vs}.

One can clearly see the exponential tails of the presented correlation functions. The exception is $K_{\rm rot}(\Vr)$ for volumetric deformation, as predicted by the theory in Eq.~(\ref{eq:Krot-r}). The length scale $ξ$ depends on the vicinity to the percolation threshold $p - p_c$, as depicted in Fig.~\ref{fig:percolation_K_vs}(e). The obtained dependence is well described by the scaling relation $ξ ∼ (p - p_c)^{-ν_{\rm na}}$ with the critical exponent $ν_{\rm na}≈0.37$, which is slightly smaller than the theoretical value $ν_{\rm na}=0.5$.

The system size of 50 unit cells in each direction is much larger than all the observed values of $ξ$. However, this does not make finite size effects negligible, since small long-range correlations proportional to $ε^2/N$ are observed for $K_{\rm div}(\Vr)$ and $K_{\rm rot}(\Vr)$. Such a background correlations are caused by the conservation of the center of mass of the finite system. They were subtracted from $K_{\rm div}(\Vr)$ and $K_{\rm rot}(\Vr)$ before being multiplied by $r$ in Fig.~\ref{fig:percolation_K_vs}.

A more detailed study of the correlation functions of nonaffine deformations in the rigidity percolation model is the subject of separate work.

\subsection{Molecular dynamics simulations: a polymer}
\label{sec:MDS}

The molecular dynamics simulations of a model atactic polystyrene in the amorphous state are performed using \textsc{lammps}~\cite{Thompson-lammps-flexible-simulation-2022}. The MARTINI force-field is applied~\cite{Beltukov-local-elastic-properties-2022, Rossi-coarsegraining-polymers-martini-2011}, which represents each styrene monomer using four coarse-grained particles. Such a force-field is well-suited for large-scale simulations of polystyrene. A total of 240 chains, each consisting of 120 monomers, were simulated. Initially, the system was generated using random placement of polymer chains, each generated by the random monomer orientation. Then the system was briefly equilibrated at a high temperature of 2000~K for 10 ps. Subsequently, the system was quenched to zero temperature at a cooling rate of 100~K/ps. Finally, local equilibrium was achieved using the FIRE minimization method~\cite{Guenole-assessment-optimization-fast-2020} ensuring that the residual forces do not exceed $10^{-9}$~kcal/(mol·nm). The simulation employed the NVT ensemble with a fixed density of $\rho=0.95$ g/cm$^3$ in a periodic simulation cell $17.0×17.0×17.0$ nm. An adaptive timestep was used during the simulation, with the maximum timestep limited to 10 fs.

Subsequently, two types of deformations were applied to the final system: volumetric and shear strains. For volumetric strain, the strain tensor is diagonal with diagonal elements $ε_{xx}=ε_{yy}=ε_{zz}=ε$. The shear deformation is represented by the traceless strain tensor~$\T{ε}$. We selected it to be diagonal with diagonal elements 
$(ε_{xx},ε_{yy},ε_{zz}) = (ε, -ε/2, -ε/2)$, allowing the use of an orthogonal simulation cell.
Additionally, two other possible permutations $(-ε/2, ε, -ε/2)$ and $(-ε/2, -ε/2, ε)$ are applied to increase the number of nonaffine displacement fields for the shear strain.
In all cases, 
a small strain $ε=10^{-5}$ is used to ensure the linear elastic regime. After the deformation of the simulation cell, the FIRE minimization method is used to minimize the energy again.

The obtained atomic displacements $\V u_i$ were mapped to the displacements on the reference lattice $\V u_j^{\rm ref}$ using Eq.~(\ref{eq:ref_map}) with the normalized smoothing matrix
\begin{align}
    ϕ_{ij} &= \tilde{ϕ}_{ij} / \sum_{i'} \tilde{ϕ}_{i'j},
    \\
    \quad \tilde{ϕ}_{ij} &= \exp\left(-\frac{(\V r_i^{\rm eq} - \V r_j^{\rm ref})^2}{2w^2}\right).
\end{align}
The smoothing parameter $w$ was varied between 0.2 and 0.6 nm. The nonaffine correlation functions have been averaged over 100 independently quenched systems. 
Due to the presence of soft spots, a known phenomenon in amorphous solids \cite{Gartner-nonlinear-plastic-modes-2016, Kapteijns-nonlinear-quasilocalized-excitations-2020, Rainone-pinching-glass-reveals-2020}, several of the generated systems have extraordinary nonaffine displacements in such regions. If the largest atomic displacement exceeded the average modulus of atomic displacements by a factor of 100, the system was excluded from the averaging process to ensure the stability of the results.

\begin{figure}[t]
    \centering
    \includegraphics[scale=0.8]{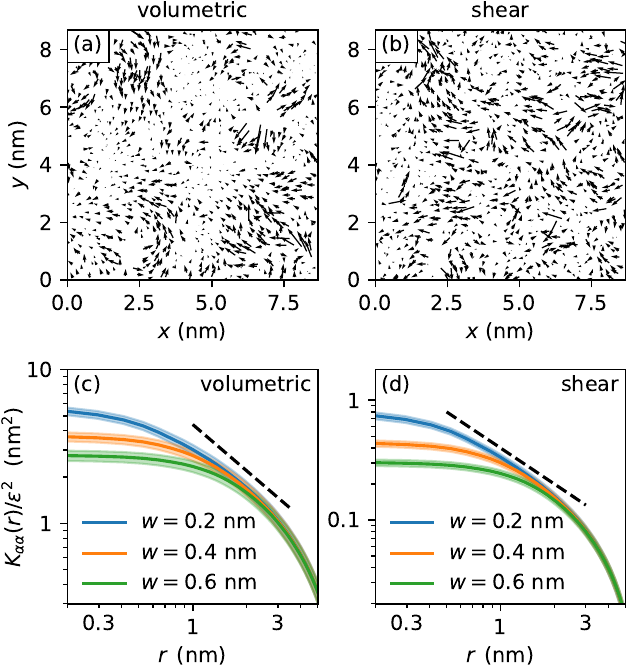}
    \caption{
    (a, b) Examples of nonaffine displacement fields in a model polystyrene under volumetric (a) and shear (b) strain. All displacements are multiplied by the factor $3·10^4$. A slice of thickness 0.5 nm in the $z$ direction is shown. In the directions $x$ and $y$ only half of the system is shown.
    (c, d) Correlation function of the nonaffine displacement field $K_{αα}(r)$ in model polystyrene system under volumetric strain (c) and shear strain (d) for different values of the smoothing parameter $w$. Shaded areas represent two standard deviations of the obtained data. Dashed lines represent the dependence $1/r$ as a visual guide.
    }
    \label{fig:dr_dr}
\end{figure}

The correlation function $K_{αα}(\V r) = ⟨u^{\rm naff}_α(\V r)\,u^{\rm naff}_α(0)⟩$ is plotted in Fig.~\ref{fig:dr_dr} for both the volumetric strain and the shear, 
along with particular examples of the nonaffine displacement fields.
For large distances $r$, one can expect the long-range power-law tail $1/r$. However, the limited system size does not allow for a clear observation of such a dependence in Fig.~\ref{fig:dr_dr}. 

\begin{figure*}[t!]
    \centering
    \includegraphics[scale=0.8]{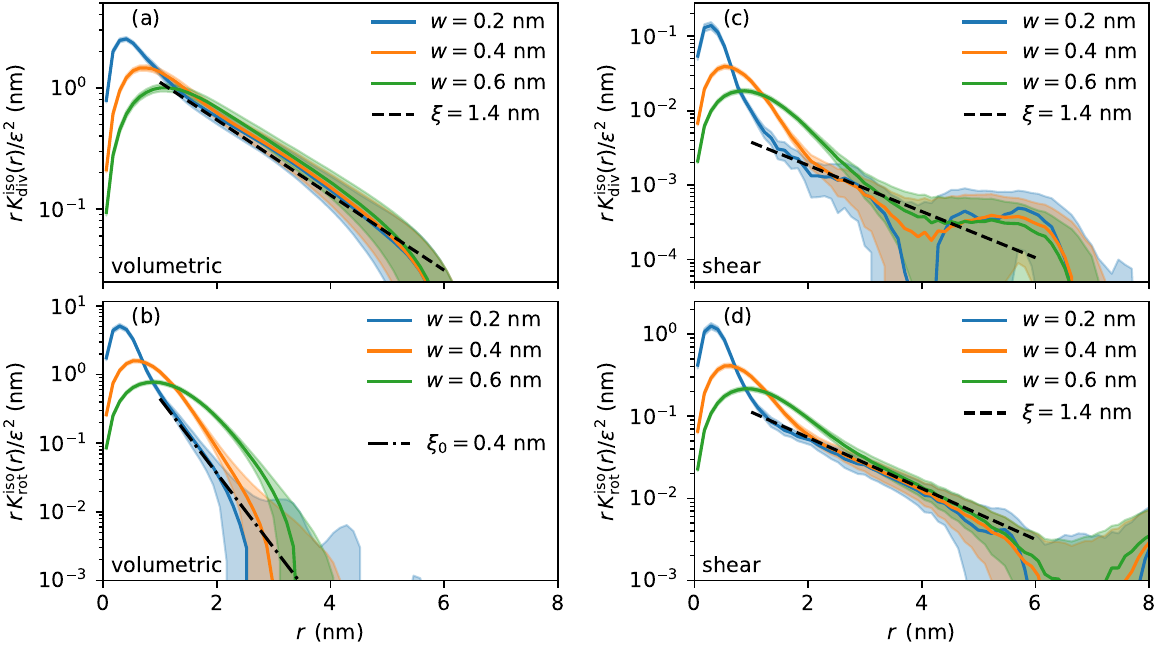}
    \caption{
    Correlation function of the divergence (a, c) and the rotor (b, d) of the nonaffine displacement field in model polystyrene under volumetric (a, b) and shear (c, d) strain for different values of the smoothing parameter $w$.  The normalization factor $r/ε^2$ is used to plot the data and the isotropic part of the correlation function is shown.  Shaded areas represent two standard deviations of the obtained data. Dashed and dash-dotted lines represent the dependencies $\exp(-r/ξ)$ and $\exp(-r/ξ_0)$, respectively.
    }
    \label{fig:K_vs}
\end{figure*}

At the same time, correlation functions $K_{\rm div}^{\rm iso}(r)$ and $K_{\rm rot}^{\rm iso}(r)$ were obtained and averaged over different directions of $\Vr$ and all the systems obtained. The normalized correlation functions $r K_{\smash{\rm div/rot}}^{\rm iso}(r)/ε^2$ are plotted in Fig.~\ref{fig:K_vs}. A small value $3·10^{-3}ε^2$ has been added to $K_{\rm rot}^{\rm iso}(r)$ to compensate finite-size effects observed in the case of the shear strain.

After some transition at distances $r<1$ nm, one can see an exponential decay ${\sim}\exp(-r/ξ)$ in Fig.~\ref{fig:K_vs}(a,c,d) up to approximately $r=6$ nm with $ξ = 1.4$ nm. Such a length scale is larger than the styrene monomer size, which is approximately $0.5$ nm. At the same time, the length scale $ξ = 1.4$ nm coincides with the length scale of the increase in the elastic moduli around nanoparticles in the amorphous polymeric matrix, associated with the suppression of the nonaffine displacements on the surface of nanoinclusions in a polymeric matrix~\cite{Beltukov-local-elastic-properties-2022, Conyuh-effective-elastic-moduli-2023}.

Using the same normalization, the correlation function for the rotor of nonaffine displacements under volumetric strain, $K_{\rm rot}^{\rm iso}(r)$, is presented in Fig.~\ref{fig:K_vs}(b). It demonstrates a much faster decay, which can be fitted by exponential decay ${\sim}\exp(-r/ξ_0)$ with a much shorter length scale $ξ_0 = 0.4$ nm, which is similar to the monomer size. Such a fast decay of $K_{\rm rot}^{\rm iso}(r)$ for volumetric strain fully corresponds to the theoretical predictions, which suggest that the term ${\sim}\exp(-r/\xi)$ with a long length scale $ξ$ is vanishing in this case.

Because the molecular dynamics simulation involves multiple types of interactions with varying strengths, a straightforward Maxwell counting approach cannot be used. Instead, the effective disorder parameter $\tk$ can be estimated from the exponential decay of the correlation functions $K_{\rm div}^{\rm iso}(r)$ and $K_{\rm rot}^{\rm iso}(r)$ as $(ξ_0/ξ)^2 \approx 0.08 \ll 1$, justifying that this system is strongly disordered.

In the present molecular dynamics study of polystyrene, a cooling rate of 100 K/ps was chosen. This relatively fast cooling enhances the degree of disorder, making the nonaffine deformations and the associated exponential decay more pronounced, which is advantageous for testing the theoretical predictions. A slower cooling rate would reduce the disorder and require significantly greater computational resources to achieve comparable statistical accuracy. A systematic study of the dependence of $ξ$ on the cooling rate for different polymer systems is an important direction for future work.

\subsection{Molecular dynamics simulations: a~Lennard-Jones glass}

To complement the numerical examples and to address the question of how the results depend on the complexity of the model system, we additionally examine a Lennard‑Jones (LJ) glass. This system is among the most thoroughly investigated amorphous systems and serves as a point of comparison with the polymer results. While the polymer system exhibits stronger quenched disorder due to chain connectivity and local packing constraints, making the exponential decay more pronounced, the LJ glass allows for more extensive computational sampling and a slower cooling rate. The combination of these two systems demonstrates the robustness of the predicted exponential correlations across different classes of amorphous materials.

The Kob-Andersen binary mixture of LJ particles is used~\cite{Kob-testing-modecoupling-theory-1995}, which contains 80\% of particles of type A and 20\% of particles of type B. In this mixture, all particles have unit masses and interact via a truncated LJ pair potential with the parameters $σ_{AA} = 1$, $σ_{BB} = 0.88$, $σ_{AB} = 0.8$, $ε_{AA} = 1$, $ε_{BB} = 0.5$, $ε_{AB} = 1.5$, and the cutoff distance $2.5$. Here and below, all the quantities are provided in the standard LJ units.

The following protocol for preparing cooled LJ systems is used. The initial system is a
simple cubic lattice consisting of $85×85×85$ LJ particles of the two types chosen randomly. The initial system is heated to the temperature $T=5$ and then cooled down at a cooling rate of $10^{-3}$ in the NVT ensemble with the fixed density $ρ=1.2$. This cooling rate lies within the typical range of $10^{-3} - 10^{-6}$ employed to study LJ glasses in the Kob-Andersen model~\cite{Sastry-signatures-distinct-dynamical-1998}. Although it is not the slowest rate in this interval, it also does not correspond to quenching as in the previous example.

The cooling process is conducted in two stages. At the first stage, the system is cooled down to the temperature $T=1$ using a timestep of 0.005. Then, the cooling process continues at the same cooling rate with a timestep of 0.01 until a near-zero temperature is reached. The second stage is repeated 5 times to increase the set of cooled systems. The final equilibrium positions are found using the FIRE minimization method. The whole preparation procedure is repeated 20 times, resulting in 100 cooled systems in local equilibrium. 

As in the previous numerical examples, two types of deformations were applied to the cooled systems: volumetric and shear strains. In both cases, the diagonal strain tensor is used with diagonal elements $(ε, ε, ε)$ for the volumetric strain and $(ε, -ε/2, -ε/2)$ together with the other two permutations for the shear strain, using a small strain magnitude of $ε=10^{-5}$. Following the deformation of the simulation cell, the FIRE minimization algorithm is used to minimize the energy again.

\begin{figure}[t!]
    \centering
    \includegraphics[scale=0.8]{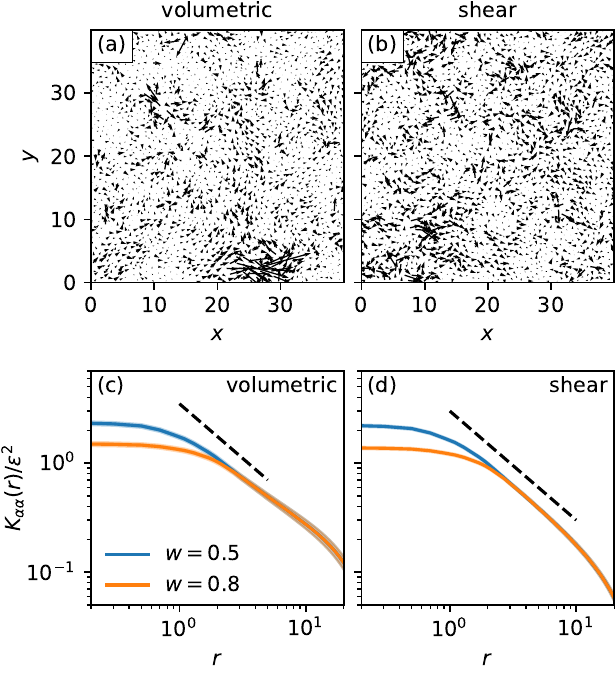}
    \caption{
    (a, b) Examples of nonaffine displacement fields in LJ glass under volumetric (a) and shear (b) strain. All displacements are multiplied by the factor $1.75·10^4$ (a) and $3.5·10^4$ (b). A slice of thickness 1 in the $z$ direction is shown. In the directions $x$ and $y$, only half of the system is shown. (c, d) 
    Correlation function of the nonaffine displacement field $K_{αα}(r)$ under volumetric (a) and shear (b) strain for different values of the smoothing parameter $w$. Shaded areas represent two standard deviations of the obtained data. Dashed lines represent the dependence $1/r$ as a visual guide.  All quantities are given in LJ units.
    }
    \label{fig:LJ_dr_dr}
\end{figure}

\begin{figure*}[t]
    \centering
    \includegraphics[scale=0.8]{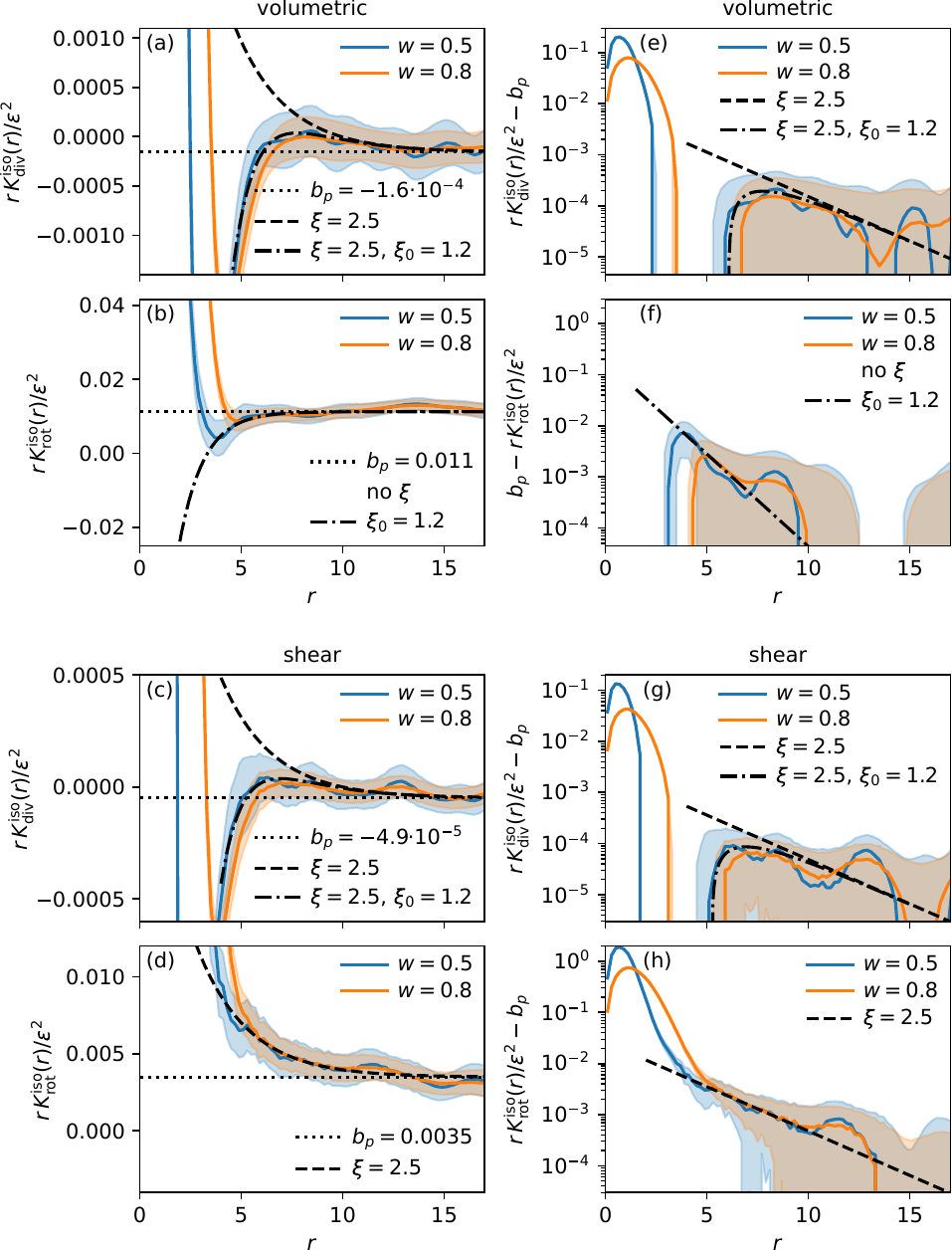}
    \caption{
    Correlation function of the divergence (a, c) and the rotor (b, d) of the nonaffine displacement field in LJ glass under volumetric (a, b) and shear (c, d) strain for different values of the smoothing parameter $w$. The normalization factor $r/ε^2$ is used to plot the data and the isotropic part of the correlation function is shown. Shaded areas represent two standard deviations of the obtained data. Dotted lines show the background value $b_p$, dashed lines show $b_p + b_ξ \exp(-r/ξ)$, dash-dotted lines show the two-exponential fit $b_p + b_ξ \exp(-r/ξ) + b_{ξ_0} \exp(-r/ξ_0)$. Panels (e–h) are the same as panels (a–d), but they are shown on a logarithmic scale with the background value $b_p$ subtracted. In addition, panel (f) is sign-inverted to highlight the negative correlations. All quantities are given in LJ units.
    }
    \label{fig:LJ_K_vs}
\end{figure*}

The correlation functions of nonaffine displacements are calculated using the same procedure as described in the previous section. The smoothing parameter $w$ was chosen to be 0.5 and 0.8. As in the previous example, several of the generated systems have extraordinary nonaffine displacements localized in soft spots. Therefore, if the largest atomic displacement exceeded the average modulus of atomic displacements by a factor of 200, the system was excluded from the averaging process to ensure the stability of the results. It excludes about 25\% of systems. A smaller value of the rejection rate does not result in a noticeable change in the correlation functions, but it increases the noise in the obtained data. A slower cooling rate is needed to decrease the number of outliers, but it requires a proportional increase in the computational time.

The correlation function $K_{αα}(\V r) = ⟨u^{\rm naff}_α(\V r)\,u^{\rm naff}_α(0)⟩$ has been averaged over different cooled systems and different directions of the vector $\Vr$. The result is plotted in Fig.~\ref{fig:LJ_dr_dr} for both the volumetric strain and the shear, along with particular examples of the nonaffine displacement fields. One can clearly see the expected long-range power-law tail $1/r$. 

The correlation functions $K_{\rm div}^{\rm iso}(r)$ and $K_{\rm rot}^{\rm iso}(r)$ have also been averaged over different cooled systems and different directions of the vector $\Vr$. Using the same normalization $r K_{\smash{\rm div/rot}}^{\rm iso}(r)/ε^2$ as in the previous Section, the correlation functions are plotted in Fig.~\ref{fig:LJ_K_vs}. The main question is whether the calculated correlation functions contain the length scale $ξ$. In contrast to previous examples, the detection of this length scale is less straightforward.

One can observe that $K_{\rm div}^{\rm iso}(r)$ has a strong short-range anticorrelation with the length scale $ξ_0≈1.2$, as shown in Fig.~\ref{fig:LJ_K_vs}(a,c). This partially hides the tail $\exp(-r/ξ)$. The length scale $ξ_0$ is comparable to the particle size and can therefore be associated with the underlying structural length scale. In particular, the typical distance between nearest-neighbor particles of type A is approximately 1.1. While the presence of the short-range anticorrelation is not unique to LJ systems (see Fig.~\ref{fig:percolation_K_vs} for the rigidity percolation), the molecular dynamics simulation requires much more computational effort to obtain well-averaged correlation functions. 

Therefore, the rotor correlation function $K_{\rm rot}^{\rm iso}(r)$ is analyzed first. The normalized correlation function $r K_{\rm rot}^{\rm iso}(r)$ has an almost constant behavior at distances $r ≳ 7$ for volumetric strain, as shown in Fig.~\ref{fig:LJ_K_vs}(b), and at distances $r ≳ 10$ for shear strain, as shown in Fig.~\ref{fig:LJ_K_vs}(d). It corresponds to the small power-law dependence $K_{\rm rot}^{\rm iso}(r) \sim b_p/r$ described by Eq.~(\ref{eq:K_rot_power}). Aside from the constant, one can clearly see the exponential behavior $\exp(-r/ξ)$ in $rK_{\rm rot}^{\rm iso}(r)$ for the shear strain with the length scale $ξ ≈ 2.5$, as shown on the logarithmic scale in Fig.~\ref{fig:LJ_K_vs}(h). For the volumetric strain, only the short-range anticorrelation is observed under the constant value $b_p$, as shown on the logarithmic scale in Fig.~\ref{fig:LJ_K_vs}(f). The corresponding length scale $ξ_0 ≈ 1.2$ is much smaller than the major length scale $ξ$.

The value of the constant $b_p$ in $r K_{\rm rot}^{\rm iso}(r)$ for the volumetric strain is 3.1 times higher than that for the shear strain. The theory given by Eq.~(\ref{eq:K_rot_power}) also predicts that these values are of the same order of magnitude. In particular, for the given volumetric and shear strains, it corresponds to a reasonable ratio of parameters $ν_2 \approx 2 ν_1$.

To quantify the normalized correlation function $r K_{\rm div}^{\rm iso}(r)/ε^2$ under volumetric and shear strain, we use the two-exponential fit $b_p + b_ξ \exp(-r/ξ) + b_0 \exp(-r/ξ_0)$ with $ξ = 2.5$ and $ξ_0 = 1.2$, see Fig.~\ref{fig:LJ_K_vs}(a,c). While the signal-to-noise ratio is small, the fit reveals the presence of the correlation length scale $ξ$ with a positive coefficient $b_ξ$. The value of $b_p$ is much smaller than that for the case of the rotor correlation function, which is in good agreement with the theory that does not predict the power-law term in the divergence correlation function. The small negative value of $b_p$ is caused by a small finite-size effect. Additional calculations show that $b_p$ shifts to negative values for all correlation functions for smaller systems.

Analogously to the polymer system, the binary Lennard-Jones mixture does not permit a straightforward definition of a bond count since distinct particle pairs interact with different strengths depending on their separation distance and particle types. Nevertheless, we can estimate the effective disorder parameter $\tk$ via the relation $(ξ_0/ξ)^2 \approx 0.23$, indicating that this system is comparatively less disordered.

As a result, the numerical simulation of LJ glass supports the results of the theory regarding the presence of a relatively large length scale $ξ$, which appears in every correlator except for $K_{\rm rot}^{\rm iso}(r)$ under volumetric strain. A small power-law contribution $K_{\rm rot}^{\rm iso}(r) \sim b_p/r$ predicted by the theory is also observed.

\section{Discussion}
\label{sec:Disc}
We consider the athermal mechanical response of an amorphous solid cooled down to zero temperature.
We study the correlation of atomic nonaffine displacements that occur under the action of a small external perturbation, so small that the system remains in a local equilibrium position and does not reach other potential energy minima. For the molecular dynamics simulations we have performed, the values of the external strain are $ε=10^{-5}$, ensuring the linear elastic regime. 

To clarify the origin and nature of the nonaffinity, we consider the correlated disorder of the 
zero-temperature
amorphous solid state by the methods of random matrix theory. The main point of the applied method is the requirement of mechanical stability of the disordered system, which corresponds to the positive semidefiniteness of the force constant matrix $\M{Φ}$ and its representation as a correlated Wishart ensemble $\M{Φ}=\M{A}·\M{A}^T$ with a correlated random matrix $\M{A}$. In amorphous solids, short-range atomic interactions typically prevail over long-range ones, resulting in a sparse covariance matrix $\FM{C}$. 
The random matrix theory remains valid as long as each particle interacts with $z$ neighboring particles and the condition $zd\gg 1$ is satisfied, as explained in Appendix~\ref{sec:aver}.

Previously, the study of the statistical properties of the correlated Wishart ensemble helped us describe the well-known vibrational phenomena of amorphous solids, such as the boson peak and the Ioffe-Regel transition during the transformation of long-wavelength acoustic phonons into the diffusion type of vibrations \cite{Beltukov-theory-sparse-random-2011,Beltukov-iofferegel-criterion-diffusion-2013, Conyuh-random-matrix-approach-2021}. The Ioffe-Regel crossover frequency and the boson peak frequency are shown to be similar, and the corresponding spatial scale, usually amounting to several nanometers, is associated with the strength of disorder and has the same order of magnitude as the heterogeneity length scale $ξ$. The classical (continuum) theory of elasticity becomes inapplicable at such scales, since it is impossible to determine a smooth dependence of a displacement on the coordinate. In other words, length scale $ξ$ separates macroscopic scales, to which the classical (continuum) theory of elasticity is applicable, and microscopic scales, on which the disorder of the system plays an essential role, which is consistent with the results of the paper \cite{Semenov-nonaffine-deformations-local-2022,Conyuh-effective-elastic-moduli-2023,Conyuh-random-matrix-approach-2021} in the framework of the random matrix theory.

Using the diagram technique outlined in Appendix~\ref{sec:aver}, we find that the spatial correlations of the nonaffine deformations $K_{αβ}(\V r) = ⟨u^{\rm naff}_α(\V r)\,u^{\rm naff}_β(0)⟩$ have two terms, $K_{αβ}(\V r) = K_{αβ}^{\rm ld}(\V r) + K_{αβ}^{\rm tw}(\V r)$. The first term demonstrates the power-law decay $K_{αβ}^{\rm ld}(\V r) \propto r^{2-d}$, which has the same spatial behavior as the Green function of an isotropic elastic medium~\cite{Mura-micromechanics-defects-solids-1987}. This result is in exact agreement with the work of DiDonna and Lubensky~\cite{DiDonna-nonaffine-correlations-random-2005}, who proposed an analytic model for correlations in systems with a random distribution of elastic moduli, and it was confirmed in the works~\cite{Maloney-correlations-elastic-response-2006, Maloney-anisotropic-power-law-2009, Mandal-singleparticle-fluctuations-directional-2013, Varnik-correlations-plasticity-sheared-2014}. The second term demonstrates the exponential decay $K_{αβ}^{\rm tw}(\V r) \propto \exp(-r/ξ)$, expressing the large-scale correlation decreasing on the heterogeneity length scale $ξ$. We suggest that this exponentially decreasing contribution was noted in \cite{Jana-correlations-nonaffine-displacements-2019, Meenakshi-characteristics-correlations-nonaffine-2022}. Since $K_{αβ}(\V r)$ includes both exponential and power-law components, it is important to note that fitting this function with only the exponential or the power-law expression might lead to inconsistent outcomes. 
Furthermore, the slowly decaying power‑law tails are strongly influenced by the finite size of the system, which can significantly complicate the analysis. Lower panels in Figs.~\ref{fig:percolation_dr_dr}, \ref{fig:dr_dr}, and \ref{fig:LJ_dr_dr} demonstrate the deviation from $1/r$ law at large distances.

Although the exponential correlation function for nonaffine displacements was derived for a correlated disorder with a specific correlation length~\cite{DiDonna-nonaffine-correlations-random-2005}, the current paper demonstrates that the length scale $ξ$ of the exponential decay depends on the disorder strength, and $ξ$ could, in theory, surpass other relevant length scales in the system, such as interatomic distance, interaction distance, and the correlation length of the disorder.

An important result of our study is the analytical expression for the divergence $K_{\rm div}(\V r) = ⟨\div \V u^{\rm naff}(\Vr)\,\div \V u^{\rm naff}(0)⟩$ and rotor $K_{\rm rot}(\V r) = ⟨\rot \V u^{\rm naff}(\V r)·\rot \V u^{\rm naff}(0)⟩$ correlation functions of the nonaffine displacement field. From a physical point of view, $K_{\rm div}(\V r)$ corresponds to the correlation between the local fluctuations of the density, and $K_{\rm rot}(\V r)$ corresponds to the correlation between local rotations.

As was shown in Section~\ref{sec:SDM}, the divergence correlation functions consist of three terms
\begin{equation}
    K_{\rm div}(\V r) = a \mkern1.5mu δ(\V r) + b_ξ \frac{e^{-r/ξ}}{r^{(d-1)/2}}. \label{eq:fit}
\end{equation}
The delta-correlated component arises from the classical power-law decay of the correlation function $K_{αβ}^{\rm ld}(\V r)$ and corresponds to white noise statistics, while the exponentially decaying contribution originates from the heterogeneity length scale $ξ$.

The correlation function $K_{\rm rot}(\V r)$ has the same structure as $K_{\rm div}(\V r)$, with an additional small power-law term
\begin{equation}
    \qquad K_{\rm rot}(\V r) = a \mkern1.5mu δ(\V r) + b_ξ \frac{e^{-r/ξ}}{r^{(d-1)/2}} + b_p r^{2-d} \label{eq:fit_rot}
\end{equation}
with some other coefficients $a$, $b_ξ$, and $b_p$ that depend on the applied stress and the strength of disorder. The only exception arises in the case of purely volumetric strain, for which the coefficient $b_ξ$ vanishes for the rotor correlation function. 

The power-law term in Eq.~(\ref{eq:fit_rot}) appears because rotation itself does not contribute to the elastic energy, which may lead to additional long-range correlations. This term arises from local spatial correlations of the force constants and was not identified in previous studies. For example, the model of Ref.~\cite{DiDonna-nonaffine-correlations-random-2005} considers correlations of the stiffness tensor but restricts the analysis to the case where all components of the corresponding correlator are governed by a single scalar function $g(\xi_0 q)$ defined at a characteristic length scale $\xi_0$. By contrast, we demonstrate that even for an isotropic amorphous solid, a complete description requires a correlation function that depends explicitly on the direction of the wavevector $\Vq$ at finite, nonzero length scales (specifically, down to the atomic scale set by $\xi_0$). This directional dependence captures the effect of local elastic anisotropy and is essential for reproducing the observed power-law tail in the rotor correlation function. Indeed, the parameters $\nu_1$ and $\nu_2$ in Eq.~(\ref{eq:K_rot_power}) have units of $(\text{pressure}/\text{length})^2$, identifying the spatial dependence of fluctuating elastic properties of amorphous solids.

The presence of the exponential tail in $K_{\rm div/rot}(\Vr)$ is confirmed by numerical investigations of the rigidity percolation model and by molecular dynamics simulations of polystyrene and the LJ glass. The fact that the rotor of the nonaffine displacement field $K_{\rm rot}(\Vr)$ under volumetric deformation does not exhibit large-scale correlations of the form $\exp(-r/\xi)$ was confirmed numerically by all three studied systems. At the same time, the presence of a small power-law tail $\sim 1/r$ at large distances in the rotor correlation function has been clearly observed in $K_{\rm rot}(\Vr)$ for the LJ glass.

For various disordered systems of arbitrary dimensionality, the heterogeneity length scale $ξ$ can be estimated from the results of molecular dynamics calculations by plotting the dependence $K_{\rm div}(\V r)$ and $K_{\rm rot}(\V r)$, while the coefficients $a$, $b_ξ$, and $b_p$ in Eqs.~(\ref{eq:fit}) and (\ref{eq:fit_rot}) can be used as fitting parameters. Molecular dynamics modeling allows for the consideration of various amorphous and polymer systems, each of which can be characterized by its own heterogeneity length scale $ξ$.

To perform molecular dynamics simulations, several methodological considerations must be taken into account. First, the system size must be sufficiently large since certain correlation functions exhibit a slow power-law decay. This behavior can generate deviations of the measured correlations from the theoretically predicted behavior, even when the simulation box size $L$ greatly exceeds the characteristic heterogeneity length scale $ξ$. Second, the applied strain should remain sufficiently small. For the system under investigation, strains exceeding $ε = 10^{-5}$ result in a distortion of the correlation functions. Furthermore, correlation functions should be averaged over a set of fully independent configurations. Quenching from a common high-temperature configuration can induce spurious long-range correlations of accidental origin. The preparation protocol and cooling rate exert a significant influence on the number of soft spots and on the density of quasilocalized modes~\cite{Lerner-effect-instantaneous-continuous-2017, Rainone-pinching-glass-reveals-2020}. In the present work, we focus on the global correlation functions of nonaffine displacements rather than on the detailed properties of individual hot spots. The theoretical framework developed here is, however, applicable to hot spots as well and predicts the emergence of the same characteristic length scale $ξ$ in their properties, while a systematic analysis of these aspects will be presented elsewhere. In the simulations reported here, we discard a small number of configurations exhibiting exceptionally large displacements in order to stabilize the averaging procedure. The resulting correlation functions show no appreciable dependence on the number of hot spots included in the averaging.

In the molecular dynamics study of polystyrene, a rapid cooling rate of 100 K/ps was chosen to enhance the degree of quenched disorder, making the nonaffine deformations and the associated exponential decay more pronounced. This makes the system particularly suitable for a proof-of-concept study. For a slower cooling rate, the magnitude of nonaffine deformations decreases, and achieving the same statistical accuracy would require computational resources beyond the scope of the present work. While a systematic investigation of the dependence of the heterogeneity length scale $ξ$ on the cooling rate is an important direction for future work, the chosen rate is sufficient to clearly observe the predicted behavior, manifesting a length scale $ξ≈1.4$ nm. Although this value is not particularly large, the calculations demonstrate the potential of such studies and pave the way for finding soft disordered materials with even larger $ξ$.

By contrast, the Lennard-Jones glass was prepared using a slower cooling rate of $10^{-3}$ (in LJ units), which is typical for such systems and allows for a more equilibrated amorphous structure. The observation of exponential correlations in both systems confirms the robustness of the predicted behavior across different preparation protocols. The results, presented in Section VI.C, are in good agreement with the theory and identify a length scale $ξ≈2.5$ (in LJ units), which is larger than the particle sizes. As a denser system with suppressed nonaffine density fluctuations, the LJ glass exhibits less prominent (but nonzero) exponential tails in $K_{\rm div}(\Vr)$ compared to $K_{\rm rot}(\Vr)$.

In the rigidity percolation model, there is no preparation protocol, and nonaffine correlation can be studied directly. In this system, the heterogeneity length scale $ξ$ depends on proximity to the percolation threshold, $p - p_c$. 
For the systems studied, the dependence is $ξ ∼ (p - p_c)^{-ν_{\rm na}}$ with the critical exponent $ν_{\rm na}≈0.37$, which differs from the prediction $ν_{\rm na}=0.5$. This difference can be attributed to the non-homogeneous distribution of over-constrained and under-constrained regions in the system, leading to a fractal behavior near the percolation threshold. The study of correlation properties of the nonaffine displacement field provides a new, direct probe for investigating criticality in rigidity percolation, which will be explored in future work. To the best of our knowledge, previous studies of the critical exponents of rigidity percolation did not examine the correlation length of the nonaffine displacement field~\cite{Arzash-shearinduced-phase-transition-2021,
Baumgarten-viscous-forces-bulk-2017,
Broedersz-criticality-isostaticity-fibre-2011}.
We did not observe a noticeable power-law behavior of $K_{\rm rot}(\Vr)$ at large distances, which is in agreement with the theory since randomly cut bonds do not have any spatial correlations.

The obtained results can also be applied to jammed solids~\cite{OHern-jamming-zero-temperature-2003, Zhang-thermal-vestige-zerotemperature-2009}. For such systems, critical behavior is observed near the isostatic point, where the number of bonds $N_{\rm b}$ is equal to the number of nontrivial degrees of freedom $N_{\rm dof}'$. This corresponds to small values of the parameter $ϰ = 1 - N_{\rm dof}'/N_{\rm b}$. Therefore, the study of the correlation properties of the nonaffine displacement field for jammed solids is of great interest. It can be compared to the proposed theory, with the expectation that the length scale $ξ ∼ ϰ^{-1/2}$ diverges near the isostatic point. The obtained scaling relation $ϰ ∼ p - p_c$ for percolation systems corresponds to the properties of jammed solids if one uses the scaling relation $ϰ ∼ z - z_c$, where $z$ is the mean connectivity of the network, and $z_c$ is the Maxwell threshold. The corresponding length $l_c ∼ (z - z_c)^{-1/2}$ associated with the jamming transition characterizes the crossover between atomistic-scale and continuum-like mechanical responses to various local and global perturbations~\cite{Lerner-breakdown-continuum-elasticity-2014, Lerner-quasilocalized-states-self-2018}. A detailed comparison of the proposed theory with numerical simulations of jammed solids can be the subject of a separate investigation.

The works~\cite{Lerner-comment-spatial-structure-2017, Lerner-quasilocalized-states-self-2018} show that the length scale $l_{\rm QLS}$ characterizing the spatial decay of the quasilocalized state of self-stress scales near the isostatic point as $l_{\rm QLS} ∼ (z-z_c)^{-1/2}$. The very same length scale observed in response functions to a local force dipole was shown in \cite{Rainone-pinching-glass-reveals-2020} to characterize the core size of soft, quasilocalized excitations in glassy matter. Therefore, the study of the elastic response to a local force dipole is of great importance for low-energy excitations in disordered systems~\cite{Lerner-lowenergy-quasilocalized-excitations-2021, Conyuh-quasilocal-vibrations-amorphous-2023}. The developed approach within the framework of the random matrix theory provides universal expressions~\eqref{eq:ld_basis} and~\eqref{eq:tw_basis} for the analysis of the elastic displacement response to both macroscopic deformation and a local force perturbation. As follows from the present studies, it can be expected that the response to a local elastic force (point or dipole) will exhibit both a power-law dependence and an exponential decay, which is also predicted by the works~\cite{Lerner-anomalous-linear-elasticity-2023, Conyuh-soft-vibrational-modes-2023}. A detailed consideration of the elastic response to a local force perturbation is the subject of further research.

The vibrational spectrum and elastic properties of glassy solids are strongly affected by internal stresses~\cite{Lerner-frustrationinduced-internal-stresses-2018}. As demonstrated in Appendix~\ref{app:pot}, taking into account such internal stresses leads to the fact that the force constant matrix $\M{Φ}$ for systems with two-body potentials can be decomposed into stabilizing and destabilizing parts:
\begin{equation}
    \M{Φ} = \M{A}·\M{A}^T - \M{B}·\M{B}^T.
\end{equation}
These two parts are strongly correlated to ensure that the resulting matrix $\M{Φ}$ is positive semidefinite and representable as $\M{Φ} = \M{A}_0·\M{A}_0^T$. Therefore, it is natural to assume that even in the case of internal stresses, near the equilibrium position $\M{Φ}$ can be represented as a correlated Wishart ensemble~\eqref{eq:AAT} with Gaussian statistics of matrix elements.

The results of this study are of great importance for the physics of nanocomposites. It was previously shown that in a highly disordered medium with rigid inclusions of nanometer sizes, an effective elastic shell is formed around such nanoinclusions due to suppression of the nonaffine displacements~\cite{Beltukov-local-elastic-properties-2022,Conyuh-effective-elastic-moduli-2023}. Elastic moduli of such a shell exceed the elastic moduli of the same material far from nanoinclusions while the thickness of the shell depends on the degree of disorder. For example, molecular dynamics simulations of polystyrene with a silica nanoparticle show an increase in stiffness at a distance of about 1.4 nm around the nanoparticle \cite{Beltukov-local-elastic-properties-2022}. 
The shell thickness $ξ$ obtained in \cite{Conyuh-effective-elastic-moduli-2023} coincides with the heterogeneity length scale $ξ$ obtained in Appendix~\ref{sec:Uncorr}.
At the same time, the additional elastic moduli of the shell decay exponentially as the distance from the nanoparticle increases. This fact highlights the relationship between local elastic properties and nonaffine deformations since the latter has an exponential correlation function $\exp(-r/ξ)$. A similar behavior of elastic moduli was also observed near the interface between amorphous and crystalline layers \cite{Semenov-nonaffine-deformations-local-2022}.

The relations obtained within the framework of the theory of random matrices can help to study the correlation properties of nonaffine deformations at a nonzero frequency $ω \neq 0$. This is especially relevant in the study of viscoelastic vibrational properties of amorphous systems. Additionally, exploring the correlation properties of nonaffine deformations in the vicinity of interfaces between media possessing distinct elastic moduli is of significant interest, especially for nanocomposite amorphous systems. This subject remains a focus for further investigation.

The generalization of the results to the case of nonzero temperatures is also of interest. At nonzero temperatures, the system fluctuates around its metastable state, experiencing slow relaxation to a more favorable metastable state. While the relaxation process is slow, the assumption of near-equilibrium behavior remains, and the main results of the paper may be applicable to nonzero temperatures. It is reasonable to assume that as temperature increases, the material becomes softer and more structurally disordered. Consequently, one might anticipate that $ξ(T)$ increases as the temperature approaches the glass transition temperature. Nonetheless, a more detailed investigation is required.

\section{Conclusion}
In this paper, the theory of correlated random matrices is applied to analyze the correlation properties of nonaffine deformations. The main contribution to the covariance matrix $\M K$ of nonaffine displacements was obtained for a near-critical system. 
The proximity to the critical point is characterized by the small control parameter $ϰ = 1 - N_{\rm dof}'/N_{\rm b}$, where $N_{\rm dof}'$ denotes the number of nontrivial (i.e., mechanically relevant) degrees of freedom and $N_{\rm b}$ is the total number of bonds. The parameter $ϰ$ can thus be interpreted as a quantitative measure of the deviation from the isostatic point. As $ϰ$ decreases, the relative fluctuations of the elastic moduli increase. Consequently, $ϰ$ serves as a measure of the degree of structural or mechanical disorder, with the limit $ϰ \to 0$ corresponding to a highly disordered system. 

In the more general situation where the number of effective bonds cannot be uniquely or unambiguously defined, the control parameter is $\,\tk = 1 - θ_0$, where $θ_0$ is the largest generalized eigenvalue quantifying the disorder. In this case, $\tk$ assumes the role of an effective distance from criticality and disorder measure, analogous to $ϰ$ in the bond-based description.

It was shown that the correlation function of the divergence of nonaffine displacements $K_{\rm div}(\Vr)$ consists of a delta-correlated component  (white noise) and exponentially decreasing large-scale correlations. The characteristic scale $ξ$, standing in the exponent, describes the heterogeneity length scale of the medium under study and diverges as $ϰ^{-1/2}$ when $ϰ\to 0$. 

The correlation function of the rotor of the nonaffine displacement field $K_{\rm rot}(\Vr)$ has also been calculated. It has the same structure as the correlation function of the divergence 
with an additional small power-law contribution at large distances. The notable case is the volumetric deformation, in which the rotor of the nonaffine displacement field lacks the exponential term $\exp(-ξ/r)$.

A numerical study of the rigidity percolation model and molecular dynamics simulations of a quenched polystyrene 
and cooled Lennard-Jones glass was 
conducted to explore the correlation properties of nonaffine displacements. The results demonstrated that the correlation of divergence and rotor of nonaffine displacements matches theoretical predictions, exhibiting exponential decay with a length scale $ξ$ greater than the typical structural length scale $\xi_0$. In the rigidity percolation model, the length scale $ξ$ diverges near the percolation threshold. It was further demonstrated that the correlation function of the rotor of the nonaffine displacement field for volumetric strain displays a vanishing of the exponential decay $\exp(-r/ξ)$, consistent with the proposed theory.

\section*{Acknowledgments}
Numerical simulations were supported by the Russian Science Foundation (grant no.\ \#22-72-10083-P). The theoretical analysis was supported by the Theoretical Physics and Mathematics Advancement Foundation ``Basis'' (grant no.\ 24-1-2-36-1).
\vspace{5mm}

\section*{Data availability}
The data that support the findings of this article are openly available~\cite{raikov_2025_17580287}.

\onecolumngrid
\appendix

\newcommand{\diag}[2][0pt]{\mathord{\raisebox{#1}{\includegraphics[page=#2,scale=0.22]{diagrams11u.pdf}}}}
\newcommand{\eqmathbox}[2][T]{\eqmakebox[#1]{$\displaystyle#2$}}

\section{Random matrix theory: the averaging procedure}
\label{sec:aver}
The averaging in the resolvent $\MG(ω) = \avgs{(\hat{Φ} - \M mω^2)^{-1}}$ can be done analytically for $\hat{Φ} = \M A\M A^T$ with a Gaussian random matrix $\M A$. In the general case, the matrix elements are correlated: $\avgs{A_{ak}A_{bl}} = \F C_{ab,kl}$. The resolvent $\MG(ω)$ can be presented as an infinite series
\begin{equation}
    \MG(ω) = \left<\frac{1}{\M A\M A^T - \M mω^2}\right> 
    = -\frac{1}{\M mω^2} - \avg{\frac{1}{\M mω^2}\M A\M A^T\frac{1}{\M mω^2}} - \avg{ \frac{1}{\M mω^2}\M A\M A^T\frac{1}{\M mω^2}\M A\M A^T\frac{1}{\M mω^2}} - \cdots
    \label{eq-app:G-series}
\end{equation}
The elements of the resolvent $\MG(ω)$ can be written explicitly in the next form:
\begin{align}
    -G_{ab}(ω) &= (\M mω^2)^{-1}_{ab} + \sum_{a_1a_2}\sum_{k_1k_2} (\M mω^2)^{-1}_{aa_1} δ^{\vphantom{-1}}_{k_1k_2} (\M mω^2)^{-1}_{b_1b} \avg{ A_{a_1k_1}A_{k_2a_2}}
    \notag\\
    &\quad + \sum_{a_1\dots a_4} \sum_{k_1\dots k_4} (\M mω^2)^{-1}_{aa_1} δ^{\vphantom{-1}}_{k_1k_2}  (\M mω^2)^{-1}_{a_2a_3} δ^{\vphantom{-1}}_{k_3k_4} (\M mω^2)^{-1}_{a_4b} \avg{A_{a_1k_1}A_{a_2k_2}A_{a_3k_3}A_{a_4k_4}}  + \cdots
    \label{eq-app:G-indices}
\end{align}
We follow from the diagram technique described in \cite{Burda-signal-noise-correlation-2004} and introduce the next graphical representation:
\begin{align*}
    (\M mω^2)^{-1}_{ab} = \diag[-7pt]{8}, \quad δ_{kl} = \diag[-7pt]{9}, \quad \avg{A_{iα,k} A_{jβ,l}} = \F C_{iαjβ, kl} = \diag[-7pt]{7}.
\end{align*}
Here the solid line joining $a$ and $b$ is the factor $(\M mω^2)^{-1}_{ab}$, the dashed line joining $k$ and $l$ is the Kronecker symbol $δ_{kl}$, and a painted joining $ak$ and $lb$ is the propagator $\F C_{ab,kl}$. Following these rules, the second term in (\ref{eq-app:G-indices}) corresponds to the next diagram:
\begin{align*}
    \sum_{a_1a_2}\sum_{k_1k_2} (\M mω^2)^{-1}_{aa_1} δ^{\vphantom{-1}}_{k_1k_2} (\M mω^2)^{-1}_{a_2b} \avg{ A_{a_1k_1}A_{a_2k_2}} = \diag[-7pt]{27}.
\end{align*}
Since the elements of the matrix $\M A$ are Gaussian random numbers, Wick's probability theorem is applicable for consecutively calculating even-point correlation functions, which are expressed as sums of all distinct products of two-point functions $\avgs{A_{a_1k_1} A_{a_2k_2}}$. Therefore, a graphical representation of the resolvent $\MG(ω)$ is
\begin{equation}
    -\left(\diag{5}\right) = 
    \diag{1} + \diag{13} + \diag{14} + \diag{15} + \diag{16} + \dots.    
    \label{eq-app:G-diagram}
\end{equation}
The presentation \eqref{eq-app:G-diagram} allows us to distinguish planar and nonplanar diagrams. For planar diagrams, the number of closed loops (closed solid line or closed dashed line) is equal to the number of double arcs. For nonplanar diagrams, the number of closed loops is less than the number of double arcs. Namely, the second diagram in \eqref{eq-app:G-diagram} is planar and contains one closed loop and one double arc, the third and fourth diagrams are planar and contain two closed loops and two double arcs, and the fifth diagram is nonplanar and contains two double arcs and only one closed loop. 

Each closed loop $Λ$ corresponds to the calculation of a trace, which gives some factor $T_Λ$ depending on the number of nonzero elements of the matrix $\M A$. If each bond involves a sufficiently large number of degrees of freedom (although the matrix $\M A$ can be a highly sparse matrix), the factor $T_Λ \gg 1$ for each closed loop $Λ$. In the case of a sufficiently filled matrix $\M A$, the factor $T_Λ \sim N$. At the condition $T_Λ \gg 1$, each planar diagram contributes much more than a nonplanar diagram with the same number of double arcs. Therefore, we can exclude nonplanar diagrams from the summation \eqref{eq-app:G-diagram} and take into account only planar diagrams.

Step-by-step solution for scalar model is described in \cite{Conyuh-effective-elastic-moduli-2023}, here we give only the final result:
\begin{equation}
    \MG(ω) = \frac{1}{\FM C:\MGs(ω) - \M m ω^2}, 
    \quad  
    \MGs(ω) = \frac{1}{\FM C^{\,T}:\MG(ω) + \M1}.
    \label{eq-app:G-closed2}
\end{equation}
Taking into account \eqref{eq-app:G-indices}, we represent the four-point resolvent \eqref{eq:tpG} as an infinite series:
\begin{align}
    \F G_{ab,a'b'} &= \avg{\left(\frac{1}{\hat{Φ} - \M mω^2}\right)_{aa'}\left(\frac{1}{\hat{Φ} - \M mω^2}\right)_{bb'}} = (\M mω^2)^{-1}_{aa'} (\M mω^2)^{-1}_{bb'} 
    \notag\\
    &\quad + (\M mω^2)^{-1}_{aa'}\sum_{b_1b_2}\sum_{k_1k_2} (\M mω^2)^{-1}_{bb_1} δ^{\vphantom{-1}}_{k_1k_2} (\M mω^2)^{-1}_{b_2b'} \avg{ A_{b_1k_1}A_{b_2k_2}}
    \notag\\
    &\quad +(\M mω^2)^{-1}_{bb'} \sum_{a_1a_2}\sum_{k_1k_2} (\M mω^2)^{-1}_{aa_1} δ^{\vphantom{-1}}_{k_1k_2} (\M mω^2)^{-1}_{aa'} \avg{ A_{a_1k_1}A_{a_2k_2}}
    \notag\\
    &\quad +\sum_{a_1a_2b_1b_2} \sum_{k_1k_2l_1l_2}(\M mω^2)^{-1}_{aa_1} δ^{\vphantom{-1}}_{k_1k_2} (\M mω^2)^{-1}_{a_2a'}(\M mω^2)^{-1}_{bb_1} δ^{\vphantom{-1}}_{l_1l_2} (\M mω^2)^{-1}_{b_2b'} \avg{A_{a_1k_1}A_{a_2k_2} A_{b_1l_1}A_{b_2l_2}}+\dots
    \label{eq-app:GG'}
\end{align}
Using the same graphical representation , we place diagrams of the first multiplier above the second one. Therefore, following Wick’s probability theorem, we can express Eq. \eqref{eq-app:GG'} in diagrams as:
\begin{equation}
    \begin{split}
        \F G_{ab,a'b'} = \diag[-22pt]{32}+ \diag[-22pt]{33} +  \diag[-22pt]{34}+\diag[-22pt]{35}+\diag[-22pt]{36}+\diag[-22pt]{37} + \dots
    \end{split}
    \label{eq-app:firstorder}
\end{equation}
In Eq. \eqref{eq-app:firstorder} can be seen that diagrams with painted joining between upper and lower ones have one closed loop less than similar ones of the ${G_{aa'}}{G_{bb'}}$. 
The diagrams that follow these two from Eq. \eqref{eq-app:firstorder} can be obtained from them in a similar way to the previous case $\MG$. It is not difficult to see that when increasing the number of connections from the upper diagrams to the lower ones, a certain pattern must be followed so that the number of closed loops is one less than in ${G_{aa'}}{G_{bb'}}$:
\begin{equation}
    \diag{28} \quad\quad \diag{29}
    \label{eq-app:secondorder}
\end{equation}
The presentation \eqref{eq-app:secondorder} allows us to distinguish stair and twisted diagrams. Note that by rearranging the indices of the lower or upper diagram, the first type can be obtained from the second, and vice versa. Therefore, we only need to keep track of one type of diagrams. 

As the structure of ladder diagrams is simpler, we will focus on this type of diagram. As can be seen from \eqref{eq-app:secondorder}, any other ladder diagram can be connected to any ladder diagram on either the left or right side and we will get a diagram of the same type. Then, considering all possible variants of the upper and lower diagrams, we get a graphical representation of the sum of all ladder diagrams:
\begin{equation}
    \diag[-16pt]{30} = \diag[-16pt]{31} + \diag[-16pt]{43}
    \label{eq-app:Ldiag}
\end{equation}
Thus, the four-point resolvent \eqref{eq:tpG} can be written as it follows:
\begin{equation}
    \F G_{ab,a'b'} = \mathcal{L}_{ab,a'b'} + \mathcal{L}_{ab',a'b}  - {G_{aa'}}{G_{b'b}},
    \label{eq-app:GG}
\end{equation}
where the term $\mathcal{L}_{ab,a'b'}$ represents the contribution of ladder diagrams, and the term $\mathcal{L}_{ab',a'b}$ represents the contribution of twisted diagrams. As it follows from the diagram \eqref{eq-app:Ldiag}, $\mathcal{L}_{ab,a'b'}$ can be written as
\begin{equation}
    \diag[-16pt]{30} = \diag[-16pt]{41} + \diag[-16pt]{44}\ ,
\end{equation}
where we introduced $\FM R$ as
\begin{equation}
    \diag[-16pt]{41} = \diag[-16pt]{31}\,, \quad \text{or} \quad \F R_{ab,a'b'} = {G_{aa'}}{G_{b'b}},
\end{equation}
and we defined $\FM T$ as
\begin{equation}
    \diag[-16pt]{39} = \diag[-15pt]{45}\,, \quad \text{or} \quad \F T_{ab,a'b'} = \sum_{klk'l'}\C_{ab,kl}\Gs_{kk'}\Gs_{ll'}\C_{a'b',k'l'}.
\end{equation}
Thus, using the diagram technique, we found that the four-point resolvent $\FM G$ is equal to
\begin{equation}
    \F G_{ab,a'b'} = \F L_{ab,a'b'} + \F L_{ab',a'b} - \F R_{ab,a'b'},
\end{equation}
where the four-point matrix $\FM L$ is defined as the solution of the equation
\begin{equation}
    \FM L = \FM R + \FM R:\FM T:\FM L,
\end{equation}
known as the two-particle Dyson equation.

\section{A model of uncorrelated bonds}
\label{sec:Uncorr}
In this Section, we provide a simple model, which represents all the properties of the strongly disordered medium discussed before. This model can be easily used to produce the corresponding force-constant matrices using random number generation.

Different bonds have different spatial positions and involve different sets of degrees of freedom. To describe the spatial correlations of nonaffine deformations in the simplest form, it is further assumed that different bonds are uncorrelated with each other, leading to a covariance of the following form \cite{Conyuh-effective-elastic-moduli-2023}:
\begin{equation}
    \C_{ab,kl} = C_{ab}^{(k)} δ_{kl}.
    \label{eq:C-uncorr}
\end{equation}
As a result, $k$-th column of the matrix $\M A$ has the Gaussian distribution with the covariance matrix $\M C^{(k)}$, which obeys sum rules (\ref{eq:C_sumrule_transl}) and (\ref{eq:C_sumrule_rot}). The nonzero elements of $\M C^{(k)}$ form a small group of nearest interacting atoms.

As it follows from the Dyson-Schwinger equations \eqref{eq:G1}--\eqref{eq:G-star}, the assumption \eqref{eq:C-uncorr} leads to the diagonal form of $\MGs$:
\begin{equation}
    \Gs_{kl} = ϰ_k δ_{kl}.
\end{equation}
For a uniform and isotropic (on average) system under consideration, $ϰ_k = ϰ$. Therefore, the solution of the Dyson-Schwinger equations \eqref{eq:G1}--\eqref{eq:G-star} in the limit $ϵ\to0$ takes the form
\begin{equation}
    \MGs = ϰ \M 1, \quad \MG = \frac{1}{ϰ} \biggl(\sum_k \M C^{(k)}\biggr)^{-1}.
\end{equation}
Since $\MG^{-1} = ϰ \sum_k \M C^{(k)}$, its Fourier transform is
\begin{equation}
    (G^{-1})_{αβ}(\V q) = \frac{ϰ}{N} \sum_{ijk} C_{iαjβ}^{(k)} e^{i\V q·(\V r_i^{\rm ref} - \V r_j^{\rm ref})}\,.
\end{equation}
For the isotropic distribution of bonds and $q\ll 1/a_0$, we obtain the usual form of Eq.~(\ref{eq:invGreen}) with Lam\'e moduli
\begin{align}
    λ &= ϰ Λ_{αβαβ}, \\
    μ &= ϰ Λ_{ααββ},
\end{align}
where the tensor
\begin{equation}
    Λ_{αβγη} = \frac{1}{N}\sum_{ijk} C_{iαjβ}^{(k)} (r_{iγ}^{\rm ref} - r_{kγ})(r_{jη}^{\rm ref} - r_{kη})
\end{equation}
represents the average properties of the bonds. 

The four-point matrix $\FM T$ has the form
\begin{equation}
    \F T_{ab,a'b'} = ϰ^2\sum_k C_{ab}^{(k)} C_{a'b'}^{(k)}.
\end{equation}
One can note that matrices $\M C^{(k)}$ are not the basis matrices of $\FM T$ orthogonal with the weight $\FM R$ since
\begin{equation}
    P_{kl} = \sum_{aba'b'} G_{ab}^\p C^{(k)}_{ba'}G_{a'b'}^\p C^{(l)}_{b'a}
    \label{eq:Pmatrix}
\end{equation}
is not equal to $N_{\rm dof}'δ_{kl}$ and does not fulfill the orthogonality criterion (\ref{eq:orthog}). However, we can apply the Fourier transform
\begin{equation}
    C_{ab}^{(\V p)} = \sum_{k} C_{ab}^{(k)} e^{i\V p·\V r_k},
\end{equation}
where $\V r_k$ is the coordinate of the center of the bond $k$. Similarly,
\begin{equation}
    P(\V p) = \frac{1}{N_{\rm b}} \sum_{kl} P_{kl} e^{i\V p·(\V r_k - \V r_l)}.
\end{equation}
Thus, the matrix $\FM T$ contains only one branch in the simple model under consideration:
\begin{equation}
    \F T_{ab,a'b'} = \frac{1}{N'_{\rm dof}}\sum_{\V p} S_{ab}^{(\V p)}θ(\V p)S_{a'b'}^{(\V p)\dag},
\end{equation}
where the basis matrices and eigenvalues are
\begin{align}
    S_{ab}^{(\V p)} &= \sqrt{\frac{1-ϰ}{P(\V p)}}C_{ab}^{(\V p)},\\
    θ(\V p) &= ϰ^2 P(\V p).
\end{align}
For small $p\ll 1/a_0$, we have the eigenfunction
\begin{equation}
    \M S^{(\V p)} = \MG^{-1}(\V p) + {\cal O}(p^2),
\end{equation}
whereas the eigenvalue is
\begin{equation}
    θ(\V p) = 1 - ϰ - ξ_0^2 p^2 + {\cal O}(p^4),
\end{equation}
where
\begin{equation}
    ξ_0^2 = \frac{ϰ^2}{2 d\thin N_{\rm b}} \sum_{kl}P_{kl} (\V r_k - \V r_l)^2.
\end{equation}
Thus, in this model of uncorrelated bonds, the approximation $\M S^{(0)} \approx \MG^{-1}$ is fulfilled and $\tl = λ$, $\tm = μ$, along with $\tk = ϰ$ describe the spatial correlations of nonaffine deformations as described by the equations in Section~\ref{sec:SDM}. The length scale $ξ = ξ_0/\sqrt{ϰ}$ coincides exactly with the one derived in \cite{Conyuh-effective-elastic-moduli-2023}.

\section{Fourier transform of nonaffine correlation function}
\label{app:fourier}
The covariance $K_{ab}$ of nonaffine displacements is expressed as the sum of two contributions, $K_{ab} = K_{ab}^{\rm ld} + K_{ab}^{\rm tw}$, given in Eq.~\eqref{eq:ld_basis} and Eq.~\eqref{eq:tw_basis}, respectively. Under uniform deformation, the covariance depends solely on the coordinate difference, as stated in Eq.~\eqref{s:fourier_K}:
\begin{align}
    K^{\rm ld}_{αβ}(\Vq) &= \frac{1}{N'_{\rm dof}N}\sum_{n\Vp}\sum_{ij} \Bigl(\MG·\M S^{(n\Vp)}·\MG\Bigr)_{i\alpha j\beta}e^{i\Vq·(\Vr_i^{\rm ref}-\Vr_j^{\rm ref})}\,\frac{θ_n(\Vp) }{1 - θ_n(\Vp)}\,\Bigl(u_{\rm aff}·\M S^{(n\Vp)\dag}·u_{\rm aff}\Bigr)\,, 
    \label{eq:Kld_app}
    \\
    K^{\rm tw}_{αβ}(\Vq) &= \frac{1}{N'_{\rm dof}N}\sum_{n\Vp}\sum_{ij} \Bigl(\MG·\M S^{(n\Vp)}·u_{\rm aff}\Bigr)_{iα}e^{i\Vq·\Vr_i^{\rm ref}}\frac{θ_n(\Vp)}{1 - θ_n(\Vp)}\,\Bigl(u_{\rm aff}·\M S^{(n\Vp)\dag}·\MG\Bigr)_{jβ}e^{-i\Vq·\Vr_j^{\rm ref}}\,,
    \label{eq:Ktw_app}
\end{align}
where the affine displacements $u^{\rm aff}_{iα} = ε_{αβ}^\p r_{jβ}^{\rm ref}$ are defined by Eq.~\eqref{eq:strain-tensor}.

Employing an arbitrary function $f(\Vp)$, the ladder term $K^{\rm ld}_{αβ}(\Vq)$ can be expressed as
\begin{align}
    \sum_{\Vp} f(\Vp) \Bigl(u_{\rm aff}·\M S^{(n\Vp)\dag}·u_{\rm aff}\Bigr) 
    &= ε_{γγ'}ε_{ηη'}\sum_{ij\Vp} r_{iγ'}^{\rm ref}r_{jη'}^{\rm ref} f(\Vp) S^{(n\Vp)*}_{iγjη}
    \\
    & = ε_{γγ'}ε_{ηη'}\sum_{ij\Vp}\frac{∂^2}{∂q_{1γ'}∂q_{2η'}}f(\Vp) S^{(n\Vp)*}_{iγjη}e^{-i\Vq_1·\Vr_i^{\rm ref} + i\Vq_2·\Vr_j^{\rm ref}}\bigg|_{\substack{\Vq_1=0\\\Vq_2=0}}
    \\
    & = N ε_{γγ'}ε_{ηη'}\frac{∂^2f(\Vq_1 - \Vq_2)S^{(n)*}_{γη}(\Vq_1,\Vq_2)}{∂q_{1γ'}∂q_{2η'}}\bigg|_{\substack{\Vq_1=0\\\Vq_2=0}}
    \\
    & = N f(0) ε_{γγ'}ε_{ηη'}\frac{∂^2S^{(n)*}_{γη}(\Vq_1,\Vq_2)}{∂q_{1γ'}∂q_{2η'}}\bigg|_{\substack{\Vq_1=0\\\Vq_2=0}}~,
\end{align}
where the constraint $\Vp = \Vq_1 - \Vq_2$, implied by Eq.~\eqref{eq:selection_p}, has been used, and the translational identity given in Eq.~(\ref{eq:sumrule_S_q1_q2}) has been applied in the final step. Consequently, the eigenvalues $\theta_n(\Vp)$ are evaluated at zero wavevector $\Vp=0$, and the remaining part of the ladder term must also be computed at $\Vp=0$:
\begin{align}
    \sum_{ij}\Bigl(\MG·\M S^{(n0)}·\MG\Bigr)_{iαjβ}e^{i\Vq (\Vr_i^{\rm ref} - \Vr_j^{\rm ref})}
    &= 
    \sum_{iji'j'} G_{iαi'α'}\,S_{i'α'j'β'}^{(n0)}\,G_{j'β'jβ} e^{i\Vq·(\Vr_i^{\rm ref} - \Vr_j^{\rm ref})}
    \\
    & = \sum_{iji'j'\Vp'} 
    G_{iαi'α'}\,e^{i\Vq·(\Vr_i^{\rm ref} - \Vr_{i'}^{\rm ref})}\, 
    S_{i'α'j'β'}^{(n\Vp')}\, e^{i\Vq·(\Vr_{i'}^{\rm ref} - \Vr_{j'}^{\rm ref})}\, 
    G_{j'β'jβ}\,e^{i\Vq·(\Vr_{j'}^{\rm ref} - \Vr_{j}^{\rm ref})}
    \\
    & = N G_{αα'}(\Vq)S^{(n)}_{α'β'}(\Vq,\Vq)G_{β'β}(\Vq),
\end{align}
where the requirement $\Vp' = 0$ due to Eq.~\eqref{eq:selection_p} has been employed. 
Using the fact that $N'_{\rm dof}/N \to d$ for a sufficiently large system, the ladder term can be expressed as
\begin{equation}
    K^{\rm ld}_{αβ}(\Vq) = \frac{1}{d}\sum_{n}G_{αα'}(\Vq)S^{(n)}_{α'β'}(\Vq,\Vq)G_{β'β}(\Vq)\,\frac{θ_n(0) }{1 - θ_n(0)}\,ε_{γγ'}ε_{ηη'}\frac{∂^2S^{(n)*}_{γη}(\Vq_1,\Vq_2)}{∂q_{1γ'}∂q_{2η'}}\bigg|_{\substack{\Vq_1=0\\\Vq_2=0}}\,.
\end{equation}

Employing an arbitrary function $g(\Vp)$, the twisted term $K^{\rm tw}_{αβ}(\Vq)$ given by Eq.~(\ref{eq:Ktw_app}) can be expressed as
\begin{align}
    \sum_{i\Vp} g(\Vp) \Bigl(\MG·\M S^{(n\Vp)}·u_{\rm aff}\Bigr)_{iα}e^{i\Vq·\Vr_i^{\rm ref}} &= \sum_{ijj'\Vp}G_{iαjγ}e^{i\Vq·(\Vr_i^{\rm ref} - \Vr_j^{\rm ref})}g(\Vp)S^{(n\Vp)}_{jγj'γ'}e^{i\Vq·\Vr_j^{\rm ref}}ε_{γ'α'}
    r_{j'α'}^{\rm ref}
    \\
    &= i G_{αγ}(\Vq)ε_{γ'α'}\sum_{jj'\Vp}\frac{∂}{∂q_{2α'}}g(\Vp)S^{(n\Vp)}_{jγj'γ'}e^{i\Vq·\Vr_j^{\rm ref}-i\Vq_2·\Vr_{j'}^{\rm ref}}\bigg|_{\Vq_2=0}
    \\
    &= iN G_{αγ}(\Vq)ε_{γ'α'}\frac{∂g(\Vq - \Vq_2)S^{(n)}_{γγ'}(\Vq,\Vq_2)}{∂q_{2α'}}\bigg|_{\Vq_2=0}
    \\
    &= iNg(\Vq) G_{αγ}(\Vq)ε_{γ'α'}\frac{∂S^{(n)}_{γγ'}(\Vq,\Vq_2)}{∂q_{2α'}}\bigg|_{\Vq_2=0}~,
\end{align}
where the translational identity given in Eq.~(\ref{eq:sumrule_S_q1_q2}) has been applied in the final step. Thus, in Eq.~\eqref{eq:Ktw_app}, the eigenvalues $\theta_n(\Vp)$ are calculated at the wavevector $\Vp=\Vq$. As shown in the main text, it leads to the appearance of the non-trivial large length scale $ξ$. For the right part of $K^{\rm tw}_{αβ}(\Vq)$ in Eq.~\eqref{eq:Ktw_app} that involves $\M S^\dagger$, the result is complex conjugated.
Finally, the twisted term is
\begin{equation}
    K^{\rm tw}_{αβ}(\Vq) = \frac{1}{d}\sum_{n}G_{αγ}(\Vq)ε_{γ'α'}\frac{∂S^{(n)}_{γγ'}(\Vq, \Vq_2)}{∂q_{2α'}} \biggr|_{\Vq_2=0}\,\frac{θ_n(\Vq)}{1 - θ_n(\Vq)} G_{βη}(\Vq)ε_{η'β'}\frac{∂S^{(n)*}_{ηη'}(\Vq, \Vq_2)}{∂q_{2β'}} \biggr|_{\Vq_2=0}\,.
\end{equation}
In the main text, the resulting expressions of $K^{\rm ld}_{αβ}(\Vq)$ and $K^{\rm tw}_{αβ}(\Vq)$ are given in Eqs.~(\ref{eq:K_ld_tw_q})–(\ref{eq:Fα_q}).

\section{Symmetry analysis of the basis matrices}
\label{app:eigenmatrices}

To establish the fundamental properties of the basis matrices $S_{iα,jβ}^\spar{n\Vp}$, it is necessary to perform an additional symmetry analysis. These basis matrices arise as solutions of the generalized eigenvalue problem (\ref{eq:eigval_TR}), formulated in terms of the four-point matrices $\FM T$ and $\FM R$:
\begin{equation}
    \FM T:\FM R:\M S^{(n\Vp)} = θ_n(\Vp) \M S^{(n\Vp)}.
    \label{eq:eigen_np}
\end{equation}
In addition to the translational and rotational sum rules of the basis matrices given by Eqs.~(\ref{eq:sumrule_S}) and (\ref{eq:rotsumrule_S}), there exists an additional significant symmetry property. We introduce the transposition superoperator (four-point matrix) $\FM F_T$, defined such that its action on a matrix $\M X$ is given by its transpose:
\begin{equation}
    \FM F_T : \M X = \M X^T.
\end{equation}
One can interchange the indices within the left and right index pairs simultaneously in the four-point matrices $\FM T$ and $\FM R$ due to their definitions given by Eqs.~(\ref{eq:R_def}) and (\ref{eq:T_def}). Therefore, the superoperator $\FM F_T$ commutes with both $\FM T$ and $\FM R$:
\begin{equation}
    \FM F_T : \FM T : \FM R = \FM T : \FM R : \FM F_T.
\end{equation}
Consequently, the matrices $\M S^{(n\Vp)}$ are also basis matrices of $\FM F_T$:
\begin{equation}
    \FM F_T : \M S^{(n\Vp)} = s_n \M S^{(n\Vp)},
    \label{eq:S_sym}
\end{equation}
where $s_n=\pm 1$, since applying $\FM F_T$ twice reproduces the initial matrix. The eigenvalues $s_n$ are therefore discrete and do not depend on the continuous parameter $\Vp$. Consequently, certain branches possess symmetric basis matrices ($s_n=1$), while others possess antisymmetric basis matrices ($s_n=-1$). We will refer to these as even and odd branches, respectively. The symmetry relation (\ref{eq:S_sym}) in the reciprocal space representation (\ref{eq:S_Fourier}) implies
\begin{equation}
    S_{αβ}^{(n)}(\Vq_1, \Vq_2) = s_n S_{βα}^{(n)}(-\Vq_2, -\Vq_1). 
    \label{eq:S_q1_q2_sym}
\end{equation}
The most important properties of the basis matrices are given by their behavior at small wavevectors $\Vq_1$ and $\Vq_2$, as it directly governs the large-scale correlation properties of nonaffine deformations.

\subsection{Even branches}

Due to the translation sum rule (\ref{eq:sumrule_S_q1_q2}), for sufficiently small wavevectors $\Vq_1$ and $\Vq_2$ we may assume the following bilinear-type expansion:
\begin{equation}
    S^{(n)}_{αβ}(\Vq_1, \Vq_2) = B_{αγβη}^{(n)}(\Vp)\,q_{1γ}^\p q_{2η}^\p,
    \label{eq:S_bilinear}
\end{equation}
where the tensor $B$ may depend on the wavevector $\Vp = \Vq_1 - \Vq_2$. This dependence arises because, for each fixed value of $\Vp$, the corresponding eigenvalue problem (\ref{eq:eigen_np}) must be solved independently. Due to the rotational rule (\ref{eq:rotsumrule_S_q1_q2}), the tensor $B$ is symmetric with respect to swapping indices within the left pair and within the right pair:
\begin{equation}
    B_{αγβη}^{(n)}(\Vp) = B_{γαβη}^{(n)}(\Vp) = B_{αγηβ}^{(n)}(\Vp) = B_{γαηβ}^{(n)}(\Vp).
\end{equation}
Due to the symmetry condition (\ref{eq:S_q1_q2_sym}) for $s_n=1$, the tensor $B$ is symmetric under the exchange of the left and right index pairs, namely:
\begin{equation}
    B_{αγβη}^{(n)}(\Vp) = B_{βηαγ}^{(n)}(\Vp).
\end{equation}
There exists an isotropic fourth-rank tensor that satisfies the symmetry constraints stated above:
\begin{equation}
    B_{αγβη}^{({\rm sym})} = \tl \, δ_{αγ}δ_{βη} + \tm \left(δ_{αη} δ_{βγ} + δ_{αβ} δ_{γη}\right),
\end{equation}
which yields
\begin{equation}
    S^{({\rm sym})}_{αβ}(\V q_1, \V q_2) = \tl \, q_{1α}^\p q_{2β}^\p + \tm \, q_{1β}^\p q_{2α}^\p + \tm \, δ_{αβ}^\p q_{1γ}^\p q_{2γ}^\p.
\end{equation}
This structure applies to the upper branch ($n=0$) and to all other non-degenerate branches. For branches that are degenerate at $\Vp=0$, the tensor $B_{αγβη}^\spar{n}(\Vp)$ can exhibit a more complicated dependence on $\Vp$ because, in this case, the secular equation must be solved for $\Vp \neq 0$.

\subsection{Odd branches}

For odd branches ($s_n=-1$), apart from the bilinear-type expansion (\ref{eq:S_bilinear}), a linear-type expansion can also be considered:
\begin{equation}
    S_{αβ}^{(n)}(\Vq_1, \Vq_2) = V_{αβγ}^{(n)}(\Vp)\,q_{1γ}^\p + W_{αβγ}^{(n)}(\Vp)\,q_{2γ}^\p,
    \label{eq:S_linear}
\end{equation}
where the tensors $V$ and $W$ may depend on the wavevector $\Vp = \Vq_1 - \Vq_2$, as in the previous case. To satisfy the translation rule given by Eq.~(\ref{eq:sumrule_S_q1_q2}), we have to require
\begin{equation}
    V_{αβγ}^{(n)}(\Vp)\,p_{γ} = W_{αβγ}^{(n)}(\Vp)\,p_{γ} = 0.
    \label{eq:odd_orthog}
\end{equation}
Using this property, we can replace $\Vq_1$ with $\Vq_1 - \Vp/2 = (\Vq_1+\Vq_2)/2$ and replace $\Vq_2$ with $\Vq_2 + \Vp/2 = (\Vq_1+\Vq_2)/2$ in Eq.~(\ref{eq:S_linear}) to obtain 
\begin{equation}
    S_{αβ}^{(n)}(\Vq_1, \Vq_2) = \frac{V_{αβγ}^{(n)}(\Vp) + W_{αβγ}^{(n)}(\Vp)}{2}(q_{1γ}^\p + q_{2γ}^\p).
\end{equation}
Therefore, without loss of generality, we can assume that $V_{αβγ}^\spar{n}(\Vp) = W_{αβγ}^\spar{n}(\Vp)$ and write
\begin{equation}
    S_{αβ}^{(n)}(\Vq_1, \Vq_2) = V_{αβγ}^{(n)}(\Vp)(q_{1γ}^\p + q_{2γ}^\p).
\end{equation}

To satisfy the rotational identity given by Eq.~(\ref{eq:rotsumrule_S_q1_q2}), we examine the derivative
\begin{equation}
    \frac{∂S_{αβ}^{(n)}(\Vq_1, \Vq_2)}{∂q_{1γ}}\bigg|_{\Vq_1 = 0} = V_{αβγ}^{(n)}(-\Vq_2) + \frac{∂V_{αβη}^{(n)}(\Vq_1 - \Vq_2)q_{2η}^\p}{∂q_{1γ}}\bigg|_{\Vq_1 = 0} =  2V_{αβγ}^{(n)}(-\Vq_2) - \frac{∂V_{αβη}^{(n)}(\Vq_1 - \Vq_2)(q_{1η}^\p - q_{2η}^\p)}{∂q_{1γ}}\bigg|_{\Vq_1 = 0}.
\end{equation}
The last term is zero due to Eq.~(\ref{eq:odd_orthog}), so we obtain the following derivatives
\begin{align}
    \frac{∂S_{αβ}^{(n)}(\Vq_1, \Vq_2)}{∂q_{1γ}}\bigg|_{\Vq_1 = 0} &= 2V_{αβγ}^{(n)}(-\Vq_2),
    \label{eq:S_odd_der1}
    \\
    \frac{∂S_{αβ}^{(n)}(\Vq_1, \Vq_2)}{∂q_{2γ}}\bigg|_{\Vq_2 = 0} &= 2V_{αβγ}^{(n)}(\Vq_1).
    \label{eq:S_odd_der2}
\end{align}
Therefore, $V_{αβγ}^\spar{n}(\Vp)$ must be symmetric over all three of its indices to satisfy the rotational identity (\ref{eq:rotsumrule_S_q1_q2}). Such a form is compatible with Eq.~(\ref{eq:S_q1_q2_sym}) for odd branches only ($s_n=-1$). The simplest symmetric form that satisfies Eq.~(\ref{eq:odd_orthog}) is
\begin{equation}
    V_{αβγ}(\Vp) = ν\,t_α t_β t_γ,
\end{equation}
where $\V t$ is a unit vector perpendicular to the wavevector $\Vp$ and $ν$ is some coefficient. We note that for small $\Vp$, the tensor $ V_{αβγ}(\Vp)$ does not depend on the magnitude of $\Vp$ since the normalization of basis matrices given by Eq.~(\ref{eq:orthog}) should be satisfied for each value of $\Vp$.

\section{Nonaffine correlation functions}
\label{app:cor}
For the general strain tensor $\T{ε}$, the nonaffine correlation function is $K_{αβ}(\Vq) = K_{αβ}^{\rm ld}(\Vq) + K_{αβ}^{\rm tw}(\Vq)$ with
\begin{align}
    K_{αβ}^{\rm ld}(\Vq) &= \frac{\tl(ε_{γγ})^2 + 2\tm ε_{γη}ε_{γη}}{d\tk} \left(\frac{\tl+2\tm}{(λ+2μ)^2}\frac{q_αq_β}{q^4} + \frac{\tm}{μ^2}\frac{q^2 δ_{αβ} - q_αq_β}{q^4}\right), \\
    K_{αβ}^{\rm tw}(\Vq) &= \frac{v_α(\Vq)v_β(\Vq)}{d(1 + ξ^2q^2)\tk}, \quad v_α(\Vq) = \frac{q_α}{λ+2μ}\left(\tl\frac{ε_{ββ}}{q^2} + 2\tm\frac{ε_{βγ}q_βq_γ}{q^4}\right) + \frac{2\tm}{μ}\left(\frac{ε_{αγ}q_γ}{q^2} - \frac{q_αε_{βγ}q_βq_γ}{q^4}\right).
\end{align}
For volumetric deformation $ε_{αβ}=εδ_{αβ}$, it simplifies to
\begin{align}
    K_{αβ}^{\rm ld}(\Vq) &= c_0 \frac{q^2δ_{αβ} - q_αq_β}{q^4} + c_1 \frac{q_αq_β}{q^4}, \\
    K_{αβ}^{\rm tw}(\Vq) &= c_2 \frac{q_αq_β}{(1 + ξ^2q^2)q^4},
\end{align}
where
\begin{equation}
    c_0 = \frac{d\tl + 2\tm}{μ^2 \tk}\tmε^2 ,
    \quad
    c_1 = \frac{(d\tl + 2\tm)(\tl + 2\tm)}{(λ + 2μ)^2\tk}ε^2,
    \quad    
    c_2 = \frac{\bigl(d\tl +2\tm)^2}{d(λ + 2μ)^2\tk}ε^2.
    \label{eq:c012}
\end{equation}
The Fourier transform (\ref{eq:Fourier}) gives the corresponding spatial correlation functions $K_{αβ}^{\rm ld}(\Vr)$ and $K_{αβ}^{\rm tw}(\Vr)$. In three dimensions ($d = 3$):
\begin{align}
     K_{αβ}^{\rm ld}(\Vr) &= \frac{(c_0 + c_1)δ_{αβ}}{8πn_{\rm at}r} + \frac{(c_0 - c_1)r_αr_β}{8πn_{\rm at}r^3},\\
     K_{αβ}^{\rm tw}(\Vr) &=\frac{c_2}{4π n_{\rm at}}\Bigg[\frac{(r^2-2ξ^2)δ_{αβ}}{2r^3} + \frac{(6ξ^2 - r^2)r_α r_β}{2r^5} + e^{-r/\xi}\left(\frac{ξ(r+ξ)δ_{αβ}}{r^3} - \frac{(r^2+3rξ+3ξ^2)r_αr_β}{r^5}\right)\Bigg].
\end{align}
In two dimensions ($d = 2$):
\begin{align}
     K_{αβ}^{\rm ld}(\Vr) &= -\frac{(c_0 + c_1)δ_{αβ}}{4πn_{\rm at}}\ln\frac{r}{r_0} + \frac{(c_0 - c_1)r_αr_β}{4πn_{\rm at}r^2},\\
     K_{αβ}^{\rm tw}(\Vr) &= -\frac{c_2}{4πn_{\rm at}}\Bigg[δ_{αβ}\left(\frac{2ξ^2}{r^2} +\ln \frac{r}{r_0} - \frac{2ξ}{r}K_1(r/ξ)\right) + \frac{r_αr_β}{r^2}\left(1-\frac{4ξ^2}{r^2}+2K_2(r/ξ)\right)\Bigg],
\end{align}
where $r_0$ is the normalization length of the order of the system size (the integral diverges for an infinite system) and $K_ν(x)$ is the modified Bessel function of the second kind, which has exponential asymptotic behavior $K_ν(x) \simeq \sqrt{π/2x}\,e^{-x}$ for large $x$.

For the general strain tensor $\T{ε}$, the correlation function of the divergence of the nonaffine displacement field defined by Eq.~(\ref{eq:Kdiv-q1}) can be written as
\begin{equation}
    K_{\rm div}(\Vq) = c_1 + \frac{c_2(\Vq)}{1 + ξ^2 q^2},
\end{equation}
where
\begin{align}
    c_1 &= \frac{\tl+2\tm}{d(λ+2μ)^2\tk}\Big(\tl(ε_{γγ})^2 + 2\tm ε_{γη}ε_{γη}\Big), \\
    c_2(\Vq) &= \frac{1}{d(λ+2μ)^2\tk} \left(\tlε_{ββ} + 2\tm \frac{q_αε_{αβ}q_β}{q^2}\right)^2.
\end{align}
For $\varepsilon_{αβ}=εδ_{αβ}$, $c_1$ and $c_2(\Vq)$ reduce to $c_1$ and $c_2$ given by Eq.~(\ref{eq:c012}), respectively. 

The correlation function of the rotor of the nonaffine displacement field defined by Eq.~(\ref{eq:Krot-q1}) can be written as
\begin{equation}
    K_{\rm rot}(\Vq) = c_3 + \frac{c_4(\Vq)}{1 + ξ^2 q^2},
\end{equation}
where
\begin{align}
    c_3 &= \frac{(d-1)\tm}{d\thin μ^2\tk}\Big(\tl(ε_{γγ})^2 + 2\tm ε_{γη}ε_{γη}\Big),  \label{eq:c1_gen}\\
    c_4(\Vq) &= \frac{4\tm^2}{d\thin μ^2\tk}\left(\frac{q_α ε_{αβ}ε_{βγ}q_γ}{q^2} - \frac{(ε_{αβ}q_α q_β)^2}{q^4}\right).  \label{eq:c3_gen}
\end{align}
For the general strain tensor $\T{ε}$, the correlation functions $K_{\rm div}(\Vq)$ and $K_{\rm rot}(\Vq)$ are anisotropic. Their isotropic parts are:
\begin{align}
    K_{\rm div}^{\rm iso}(\Vq) &= c_1 + \frac{c_2}{1 + ξ^2 q^2},   \\
    K_{\rm rot}^{\rm iso}(\Vq) &= c_3 + \frac{c_4}{1 + ξ^2 q^2},  
\end{align}
where $c_1$ and $c_3$ are defined in Eqs.~(\ref{eq:c1_gen}) and (\ref{eq:c3_gen}), while $c_2$ and $c_4$ are the isotropic parts of $c_2(\Vq)$ and $c_4(\Vq)$, respectively:
\begin{align}
    c_2 &= \frac{1}{d^3(λ+2μ)^2\tk}\left((d\tl +2\tm)^2(ε_{αα})^2 + \frac{8\tm^2(d\thin ε_{αβ}ε_{βα}-(ε_{αα})^2)}{d+2}\right),  \label{eq:c2_iso}\\
    c_4 &= \frac{4\tm^2}{d^2(d+2)μ^2\tk}\Big(d\thin ε_{αβ}ε_{βα} - (ε_{αα})^2\Big).  \label{eq:c4_iso}
\end{align}
The inverse Fourier transform (\ref{eq:Fourier}) gives the corresponding spatial correlation functions $K_{\rm div}^{\rm iso}(\Vr)$ and $K_{\rm rot}^{\rm iso}(\Vr)$:
\begin{align}
    K_{\rm div}^{\rm iso}(\Vr) &= \frac{c_1}{n_{\rm at}}δ(\Vr) + \frac{c_2}{n_{\rm at}}D(\Vr),   \\
    K_{\rm rot}^{\rm iso}(\Vr) &= \frac{c_3
    }{n_{\rm at}}δ(\Vr) + \frac{c_4}{n_{\rm at}}D(\Vr).  
\end{align}

The expressions of $K_{\rm div}(\Vq)$ and $K_{\rm rot}(\Vq)$ as well as $K_{\rm div}(\Vr)$ and $K_{\rm rot}(\Vr)$ for the particular case of the volumetric deformation are given in the main text in Section \ref{sec:SDM}.

\section{Force-constant matrix for two-body potentials}
\label{app:pot}

The force constant matrix $\M{Φ}$ for systems with two-body potentials can be written explicitly. Such systems have the potential energy of the form $U = \frac{1}{2} \sum_{ij} U_{ij}(r_{ij})$. In this case, the force-constant matrix can be decomposed into longitudinal and transverse contributions~\cite{Wyart-rigidity-amorphous-solids-2005, Lerner-frustrationinduced-internal-stresses-2018}:
\begin{align}
    Φ_{kα,lβ} &= Φ_{kα,lβ}^{\rm long} + Φ_{kα,lβ}^{\rm trans},
    \\[0.2em]
    Φ_{kα,lβ}^{\rm long} &= \sum_{ij} k_{ij} \frac{∂r_{ij}}{∂r_{iα}} \frac{∂r_{ij}}{∂r_{iβ}},
    \\
    Φ_{kα,lβ}^{\rm trans} &= -\sum_{ij}f_{ij}\frac{∂^2r_{ij}}{∂r_{iα}∂r_{iβ}},
\end{align}
where $r_{ij}$ is the distance between the particles $i$ and $j$, $k_{ij} = U''_{ij}(r_{ij})$ is the stiffness, and $f_{ij} = -U'_{ij}(r_{ij})$ is the internal force (stress) within this pair. Both contributions can be written using the normal vector $\V{n}_{ij} = \V{r}_{ij}/r_{ij}$ and $d-1$ tangential vectors $\V{t}_{ij}^{(s)}$ that are orthogonal to the normal vector $\V{n}_{ij}$ (in three dimensions, there are two tangential vectors marked by index $s=1,2$):
\begin{align}
    Φ_{kα,lβ}^{\rm long} &= \sum_{ij} k_{ij} n_{ijα} n_{ijβ} (δ_{jk} - δ_{ik})(δ_{jl} - δ_{il}),   \label{eq:Phi_long}
    \\
    Φ_{kα,lβ}^{\rm trans} &= -\sum_{ij}\sum_{s=1}^{d-1}\frac{f_{ij}}{r_{ij}} t_{ijα}^{(s)}t_{ijβ}^{(s)}(δ_{jk} - δ_{ik})(δ_{jl} - δ_{il}).   \label{eq:Phi_trans}
\end{align}
For an unstressed system with harmonic interactions, $k_{ij} ≥ 0$ and $f_{ij} = 0$, which result in positive-semidefinite terms in Eq.~(\ref{eq:Phi_long}). In this case, one can write the force-constant matrix directly as $\M{Φ} = \M{A}\M{A}^T$. However, for the stressed system, the terms with $f_{ij} > 0$ are negative-semidefinite. In the general case, some coefficients $k_{ij}$ can be negative (like those for the Lennard-Jones potential at large distances), which gives negative-semidefinite terms in Eq.~(\ref{eq:Phi_long}). Thus, both contributions may have stable terms ($k_{ij} > 0$ and $f_{ij} < 0$) as well as unstable terms ($k_{ij} < 0$ and $f_{ij} > 0$). However, stable and unstable terms are correlated in a complicated way that ensures the stability of the system under consideration.

The longitudinal and transverse force-constant matrices can be decomposed into stabilizing and destabilizing parts as:
\begin{align}
    \M{Φ}^{\rm long} &= \M{A}_{\rm long}·\M{A}_{\rm long}^T - \M{B}_{\rm long}·\M{B}_{\rm long}^T,
    \\
    \M{Φ}^{\rm trans} &= \M{A}_{\rm trans}·\M{A}_{\rm trans}^T - \M{B}_{\rm trans}·\M{B}_{\rm trans}^T,
\end{align}
where
\begin{align}
    A_{kα,ij}^{\rm long} &= 
    \begin{cases}
        \sqrt{k_{ij}} (δ_{jk} - δ_{ik}) n_{ijα}, & k_{ij} > 0,
        \\
        0, & \text{elsewhere},
    \end{cases}
    \\
    B_{kα,ij}^{\rm long} &= 
    \begin{cases}
        \sqrt{-k_{ij}} (δ_{jk} - δ_{ik}) n_{ijα}, & k_{ij} < 0,
        \\
        0, & \text{elsewhere},
    \end{cases}
\end{align}
\begin{align}
    A_{kα,ijs}^{\rm trans} &= 
    \begin{cases}
        \sqrt{-\frac{f_{ij}}{r_{ij}}} (δ_{jk} - δ_{ik}) t_{ijα}^{(s)}, & f_{ij} < 0,
        \\
        0, & \text{elsewhere},
    \end{cases}
    \\
    B_{kα,ijs}^{\rm trans} &= 
    \begin{cases}
        \sqrt{\frac{f_{ij}}{r_{ij}}} (δ_{jk} - δ_{ik}) t_{ijα}^{(s)}, & f_{ij} > 0,
        \\
        0, & \text{elsewhere}.
    \end{cases}
\end{align}
Using the horizontal stacking of the matrices
\begin{align}
    \M{A} &= (\M{A}^{\rm long},\M{A}^{\rm trans}), \\
    \M{B} &= (\M{B}^{\rm long},\M{B}^{\rm trans}),
\end{align}
the final force-constant matrix $\M{Φ}$ can be written as
\begin{equation}
    \M{Φ} = \M{A}·\M{A}^T - \M{B}·\M{B}^T.
\end{equation}
Thus, we can explicitly write the stabilizing (first) and destabilizing (second) parts of the force-constant matrix in the case of a pair potential. 

However, the matrices $\M{A}$ and $\M{B}$ are strongly correlated with each other to ensure that the resulting matrix $\M{Φ}$ is positive-semidefinite. Therefore, the force-constant matrix $\M{Φ}$ can always be represented as 
\begin{equation}
    \M{Φ} = \M{A}_0·\M{A}_0^T.
\end{equation}
The matrix $\M{A}_0$ can be explicitly expressed through the matrices $\M{A}$ and $\M{B}$ as follows~\cite{Conyuh-soft-vibrational-modes-2023}:
\begin{align}
    \M{A}_0 &= \M{A} - \M{B}·\M{W}^{-1}·\M{B}^T·\big(\M{A}·\M{A}^T\big)^{-1}·\M{A},\\
    \M{W} &= \M{I} + \sqrt{\M{I} - \M{B}^T·\big(\M{A}·\M{A}^T\big)^{-1}·\M{B}}.
\end{align}
This representation is possible if
\begin{equation}
    \big\lVert \M{B}^T·(\M{A}·\M{A}^T)^{-1}·\M{B}\big\rVert_2 \leq 1,  \label{eq:cond}
\end{equation}
which means that for a stable system the maximum eigenvalue of the matrix $\M{B}^T·(\M{A}·\M{A}^T)^{-1}·\M{B}$ is no greater than one. One can show that condition (\ref{eq:cond}) is satisfied for any positive-semidefinite matrix $\M{Φ} = \M{A}·\M{A}^T - \M{B}·\M{B}^T$ while $\M{A}·\M{A}^T$ is invertible.

The stability criterion is satisfied if the contribution of unstable bonds is not large enough. In this case, the elastic response is described by the force constant matrix of $\M{Φ} = \M{A}_0·\M{A}_0^T$. If the stability criterion is violated, the system becomes unstable near the equilibrium position under consideration. In this case, the system must reconstruct and occupy a new equilibrium position determined by the new force constant matrix. Consequently, the assumption that the matrix $\M{A}_0$ has correlated Gaussian random entries is a more realistic assumption than assigning such statistics directly to the matrices $\M{A}$ and $\M{B}$.

\twocolumngrid
\bibliography{refs1.bib}

\end{document}